\documentclass[twocolumn,times]{aastex631}

\usepackage{amsmath}
\usepackage{mathtools}
\usepackage{eucal}
\usepackage{amssymb}
\usepackage{enumitem}
\usepackage{svg}
\usepackage{comment}

\definecolor{webgreen}{rgb}{0,.5,0}
\definecolor{webbrown}{rgb}{.6,0,0}

\def \beq{\begin{equation}}
\def \eeq{\end{equation}}
\def \bea{\begin{eqnarray}}
\def \eea{\end{eqnarray}}

\received{\today}

\newcommand{\TKep}{\mbox{$ T_{\rm Kep} $}}

\newcommand{\Tlib}{\mbox{$T_{\rm lib}$}} 
\newcommand{\Tbb}{T_{\rm 2b}}
\newcommand{\Tl}{T_{\ell}} 
\newcommand{\Tsec}{T_{\rm sec}}
\newcommand{\Tlz}{T_{z}}
\newcommand{\Tlc}{T_{\rm lc}}
\newcommand{\Tgrow}{T_{\rm grow}}
\newcommand{\Tdecay}{T_{\rm decay}} 
\newcommand{\vesc}{v_{\rm esc}} 

\newcommand{\Mbh}{M_{\bullet}}
\newcommand{\lnL}{\ln{\Lambda}}
\newcommand{\rtid}{r_{\rm tid}}

\newcommand{\p}{\partial}
\newcommand{\rmd}{\mbox{${\rm d}$}}

\newcommand{\Msun}{\mbox{$ M_\odot $  }}
\newcommand{\Rsun}{R_{\odot}}

\newcommand{\Hs}{\mbox{$H_{\rm s}$}}

\newcommand{\Md}{\mbox{$M_{\rm d}$}}
\newcommand{\rh}{\mbox{$r_{\rm h}$}}

\newcommand{\vrel}{v_{\rm rel}} 
\newcommand{\Cdf}{C_{\rm df}}

\newcommand{\vg}{v_{\rm g}} 
\newcommand{\vstar}{v_{\star}} 

\newcommand{\rhoaxis}{\rho_{\rm a}} 
\newcommand{\Phiaxis}{\Phi_{\rm a}} 
\newcommand{\gaxis}{\gamma_{\rm a}} 
\newcommand{\rhod}{\rho_{\rm d}} 
\newcommand{\phid}{\varphi_{\rm d}}
\newcommand{\Phid}{\Phi_{\rm d}} 
\newcommand{\Sigd}{\Sigma_{\rm d}}

\newcommand{\rhoc}{\rho_{\star}} 
\newcommand{\phic}{\varphi_{\star}}
\newcommand{\Phic}{\Phi_{\star}} 
\newcommand{\Mstar}{M_{\star}}

\newcommand{\eLC}{e_{\rm lc}}
\newcommand{\llc}{\ell_{\rm lc}} 

\newcommand{\mstar}{m_{\star}}
\newcommand{\Rstar}{R_{\star}}

\newcommand{\jcrit}{j_{\rm crit}}
\newcommand{\lhyp}{\ell_{\rm hyp}}

\newcommand{\gtde}{g_{\rm tde}}

\newcommand{\algam}{\alpha_{\gamma}} 
\newcommand{\mathK}{\mathcal{K}}
\newcommand{\mathF}{\mathcal{F}} 
\newcommand{\mathE}{\mathcal{E}}

\newcommand{\jc}{j_{\rm c}}
\newcommand{\jl}{j_{\rm l}}
\newcommand{\js}{j_{\rm s}} 

\newcommand{\jco}{j_{\rm c 0}}

\newcommand{\muc}{\mu_{\rm c}} 

\newcommand{\vplus}{v_{+}}
\newcommand{\vminus}{v_{-}} 
\newcommand{\mue}{\mu_{\rm e}} 

\newcommand{\lmin}{\ell_{\rm min}}
\newcommand{\lmax}{\ell_{\rm max}} 
\newcommand{\gmin}{g_{\rm min}}
\newcommand{\gmax}{g_{\rm max}} 
\newcommand{\Ntde}{N_{\rm tde}} 

\newcommand{\NdotCLav}{ \langle \dot{N}_{\rm cl} \rangle }
\newcommand{\NdotCL}{  \dot{N}_{\rm cl}  }

\newcommand{\NdotBB}{  \dot{N}_{\rm 2b}  }

\newcommand{\alpaxi}{\alpha_{\rm axi}}
\newcommand{\Reff}{\mathcal{R}_{\rm eff}} 
\newcommand{\qs}{q_{\rm s}}
\newcommand{\qaxi}{q_{\rm axi}}

\newcommand{\EA}{E$+$A}

\newcommand{\lz}{\ell_z} 

\newcommand{\Fdis}{F_{\rm dis}}
\newcommand{\rcap}{\hat{r}}
\newcommand{\zcap}{\hat{z}}
\newcommand{\phicap}{\hat{\phi}} 

\newcommand{\Fdisvec}{\Vec{F}_{\rm dis}} 
\newcommand{\vrelvec}{\Vec{v}_{\rm rel}} 
\newcommand{\vgvec}{\Vec{v}_{\rm g}} 
\newcommand{\vstarvec}{\Vec{v}_{\star}} 
\newcommand{\vgr}{v_{r}^{g}}
\newcommand{\vgphip}{v_{\phi p}^{g}} 
\newcommand{\vgzp}{v_{z p}^{g}} 
\newcommand{\vstarr}{v_{r}^{\star}} 
\newcommand{\vstarphip}{v_{\phi p}^{\star}} 
\newcommand{\vstarzp}{v_{z p}^{\star}} 
\newcommand{\vstarz}{v_{z}^{\star}}

\newcommand{\lell}{\ell_{\rm ell}}

\newcommand{\fagn}{f_{\rm agn}}

\newcommand{\Nbbo}{N_{\rm 2b}^{\rm (o)}}
\newcommand{\Nclo}{N_{\rm cl}^{\rm (o)}}

\begin{document}

\title{Elevated Rates of Tidal Disruption Events in Active Galactic Nuclei}

\correspondingauthor{Karamveer Kaur}
\email{karamveerkaur30@gmail.com}

\author[0000-0001-6474-4402]{Karamveer Kaur}
\affiliation{Technion - Israel Institute of Technology, Haifa, 3200002, Israel}
\affiliation{Racah Institute of Physics, The Hebrew University of Jerusalem, 9190401, Israel}

\author[0000-0002-4337-9458]{Nicholas C. Stone}
\affiliation{Racah Institute of Physics, The Hebrew University of Jerusalem, 9190401, Israel}

\shorttitle{TDE Rates in AGN}
\shortauthors{Kaur and Stone}

\begin{abstract}

Advances in time domain astronomy have produced a growing population of flares from galactic nuclei, including both tidal disruption events (TDEs) and flares in active galactic nuclei (AGN).  Because TDEs are uncommon and AGN variability is abundant, large-amplitude AGN flares are usually not categorized as TDEs. While TDEs are normally channelled by the collisional process of two-body scatterings  over relaxation timescale, the quadrupole moment of a gas disk alters the stellar orbits, allowing them to collisionlessly approach the central massive black hole (MBH). This leads to an effectively enlarged loss cone, the \emph{loss wedge}. Earlier studies found a moderate enhancement, up to a factor $\sim 2-3$, of TDE rates $\NdotBB$ for a static axisymmetric perturbation. Here we study the loss wedge problem for an evolving AGN disk, which can capture large number of stars into the growing loss wedge over much shorter times. The rates $\NdotCL$ of collisionless TDEs produced by these time-evolving disks are much higher than the collisional rates $\NdotBB$ in a static loss wedge. We calculate the response of a stellar population to the axisymmetric potential of an adiabatically growing AGN disk and find that the highest rates of collisionless TDEs are achieved for the largest (i) MBH masses $\Mbh$ and (ii) disk masses $\Md$. For $\Mbh\sim 10^7 \Msun$ and $\Md \sim 0.1 \Mbh$, the rate enhancement can be up to a factor $\NdotCL/\NdotBB \sim 10$. The orbits of collisionless TDEs sometimes have a preferred orientation in apses, carrying implications for observational signatures of resulting flares.

\end{abstract}


\section{Introduction} 

The centers of many galaxies are home to dense nuclear star clusters (NSCs) hosting a central massive black hole (MBH;  \citealt{Kormendy_Ho_2013,Neumayer_2020}). Diverse dynamical phenomena in these dense environments may channel a star onto a high eccentricity orbit, leading to its disruption when it encounters the strong tidal gravity of MBH \citep{Hills_1975, Rees88, Phinney_1989}. These tidal disruption events (TDEs) are observed as multi-wavelength nuclear flares over timescales spanning a few months to years \citep{Bade_1996,Komossa_1999,Gezari_2006,Gezari_2008,van_Velzen_2011, vanVelzen+19}. Over the last several years, advances in time-domain surveys have increased the number of TDE candidates to roughly 100 \citep{Auchettl_2017,van_Velzen2020,Gezari_2021,Sazonov_2021,Lin_2022,Goldtooth_2023,Hammerstein_2023,Yao_2023}. These numbers are expected to soar to the thousands in the near future due to upcoming wide-field facilities, like \emph{LSST} and {\it ULTRASAT} \citep{Bricman_2020,Hambleton_2023,Shvartzvald_2023}.

Observationally inferred per-galaxy TDE rates are generally $\dot{N}\sim 10^{-5}-10^{-4}~{\rm yr}^{-1}$ \citep{Esquej_2008,Gezari_2009,van_Velzen_Farrar_2014,vanVelzen_2018,Lin_2022,Yao_2023,Masterson_2024}.  Theoretical rates predicted from dynamical modeling are systematically higher, usually by a factor of a few to an order of magnitude (\citealt{Syer_Ulmer_1999,Magorrian_Tremaine_1999,Wang_Merritt_2004,Stone_Metzer_2016,Pfister_2021,Broggi_2022}; see \citealt{Stone_2020} for a review, though see also \citealt{Teboul_2024, Polkas_2023}). Further complicating the picture, the host galaxies of optical and X-ray TDEs preferentially belong to a rare sub-group of post-starburst galaxies, the so-called \EA \, galaxies identifiable from strong Balmer absorption lines \citep{Arcavi_2014,French_2016,Law-Smith_2017,Graur_2018,Hammerstein_2021}. This observed host preference further enhances the rate discrepancy, though the discovery of IR-selected TDEs in the dust-enshrouded nuclei of star-forming galaxies may alleviate this problem to some extent \citep{Wang_2022_TDEhost_star_forming,Reynolds_2022,Onori_2022,Masterson_2024}. 

The observational characteristics of TDEs are diverse \citep{Kajava_2020,Malyali_2023b,Hammerstein_2023,Yao_2023} and depend on the multi-dimensional parameter space associated with properties of the star, MBH, stellar orbit, nuclear environment and orientation of the line of sight \citep{Lodato_2009,GuillochonRamirezRuiz13,Guillochon_2015, Hayasaki+16,Generozov+17, Dai_2018, Chan_Piran2021}. This makes it hard to identify TDE-like flares, especially in complex galactic environments, like active galactic nuclei (AGNs) which have a pre-existing accretion disk \citep{Zabludoff_2021,Saha_2023,Brogan_2023}.  
Hence, %TDEs in AGNs have been traditionally ignored, while 
TDE-like flares in AGN are usually attributed to high-amplitude AGN variability or disk instabilities \citep{Brandt_1995,Shappee_2014,Trakhtenbrot2019} and discarded to avoid sample contamination. However, there are a few recently discovered transients that can be described as TDEs occurring in an AGN \citep{Blanchard17, Tadhunter_2017,Cannizzaro_2022,Homan_2023,Goodwin_2024,Charalampopoulos_2024,Zhang_2024,Hinkle_2024}. AGN TDEs are of particular interest because of the role that high-amplitude AGN flares (possibly generated by TDEs) may play in the origins of ultra-high energy cosmic rays \citep{Farrar09,Farrar_2014} and IceCube-detected neutrinos \citep{Reusch22, vanVelzen24}\footnote{In principle, vacuum TDEs may be capable of producing these high energy particles on their own \citep{Stein21, Cannizzaro21, Piran23}, but if particle acceleration occurs in relativistic jets, a pre-existing AGN disk may be necessary. Vacuum TDEs begin with minimal net magnetic flux (\citealt{Tchekhovskoy14}; a potentially important ingredient in jet production), so the large magnetic flux supplied by an AGN disk \citep{Kelley14}  be a necessary ingredient in powering TDE jets.}.

TDEs in vacuum galactic nuclei are driven by weak two-body (2B) scatterings of stars, which occur over a timescale $\Tbb$. A star (of mass $\mstar$ and radius $\Rstar$) is tidally disrupted when its periapsis $r_{\rm p}$ becomes smaller than the tidal radius $\rtid = (\Mbh/\mstar)^{1/3} \Rstar$. This condition on $r_{\rm p}$ %(or on angular momentum $L$) 
is equivalent, in spherical symmetry, to requiring that the stellar velocity vector falls into the so-called ``loss cone'' \citep{FrankRees76, LightmanShapiro77, CohnKulsrud78} in velocity space.  Near the MBH, this region  is empty, and the rate of TDEs is set by the rate at which 2B scatterings repopulate the empty loss cone.  Both the TDE rate discrepancy and the recent discoveries of TDEs in AGNs or dust-enshrouded, gas-rich nuclei motivate careful consideration of TDE rates in the complex environments of realistic galactic nuclei. Some important complications include central MBH binaries interacting with the surrounding star cluster \citep{Ivanov_2005,Chen_2009,Chen_2011,Wegg_Bode_2011}; star clusters with very high stellar densities and/or radially biased velocity anisotropies \citep{Stone_Metzer_2016,Stone_vanVelzen_2016,Stone_2018}; post-merger eccentric stellar disks \citep{Madigan_2018}; and a dynamically cold stellar disk arising after recent star formation in a progenitor AGN \citep{Kaur_2018disk,Wang_2024}.    

In contrast to the aforementioned TDE rate calculations based on collisional\footnote{``Collisional'' does not refer here to physical collisions between stars, but rather to 2B scatterings represented by a collision operator in the Boltzmann equation.} 2B processes, there are other loss cone refilling mechanisms based on collisionless stellar orbits created by asphericity in the central star cluster \citep{Magorrian_Tremaine_1999,Merritt_Vasiliev_2011,Vasiliev_Merritt2013}, or by the presence of a massive axisymmetric gas disk \citep{Karas_Subr_2007}. In particular, the orbit-averaged (or secular) dynamics of axisymmetric systems creates a family of librating orbits (also called saucer orbits) that can attain very high eccentricities $e$ over a time $\Tlib$ during a part of their libration cycle \citep{Sambhus2000,Vasiliev_Merritt2013}. This effectively extends the loss cone into a \emph{loss wedge}, such that orbits with low initial eccentricity can enter the loss cone over a libration timescale $\Tlib$. Past works investigated the relaxation-driven enhancement of TDE rates due to collisional filling of a loss wedge over the relaxation timescale $\Tbb$, as a result of 2B scatterings \citep{Magorrian_Tremaine_1999,Vasiliev_Merritt2013}. 

In this work, we consider a time-dependent scenario describing a growing AGN gas disk, such that the disk growth timescale $\Tgrow$ is much longer than $\Tlib$. Fresh stellar orbits are efficiently captured into a growing loss wedge over the timescale $\Tgrow$, and can evolve into a \emph{collisionless} TDE owing to their secular libration over $\Tlib$. This capture process may lead to 
an even larger enhancement of TDE rates than has been previously considered in aspherical potentials.  We demonstrate this with a simple model of an adiabatically growing gas disk, for which we can evaluate the non-linear deformation of the star cluster's distribution function (DF) analytically \citep{Sridhar_1996}. This leads us to a straightforward evaluation of the time-dependent collisionless TDE rate in a growing AGN.

In \S~\ref{sec_phy_setup}, we describe the physical setup of the model NSC and gas disk, their secular dynamics, and the dynamical origins of a collisionless TDE. In \S~\ref{sec_CL_calci}, we evaluate the DF deformation of the NSC as a response to the adiabatically growing gas disk. We evaluate collisionless TDE rates and compare them with the standard picture of collisional TDE rates in axisymmetric systems. Finally, we conclude in \S~\ref{sec_discus}.

\section{Secular dynamics of cusp-disk system}
\label{sec_phy_setup} 

We consider an NSC with a central MBH of mass $\Mbh$ and an axisymmetric time-evolving gas disk of mass $\Md(t)$ inside the MBH radius of influence, $\rh$. Initially, the NSC is spherically symmetric with a power-law density cusp, with a general anisotropic distribution function (DF) $F_0$. Secular dynamics, which is orbit-averaged over mean motion, serves as a convenient and valid tool for a dynamical description of the system within $\rh$. As the disk gradually grows in mass, the secular orbits evolve from circulating rosettes (with fixed eccentricity $e$) in a spherical geometry to the two families of circulating and librating orbits (with evolving $e$) in an axisymmetric geometry.

For radii $r<\rh$, the potential is nearly Keplerian and the phase space of stellar orbits can be defined by Delaunay variables with actions $\{ I,L,L_z \}$ \citep{Murray_Dermott1999book,Sambhus2000}. $I = \sqrt{G \Mbh a}$ has a one-to-one relationship with the semi-major axis $a$; $L = I \sqrt{1-e^2}$ is the total angular momentum; $L_z = L \cos{i}$ is the $z$-component of angular momentum, with $i$ being orbital inclination. The canonically conjugate angles are $\{ w,g,h\}$, with $w$ the mean anomaly, $g$ the argument of periapsis, and $h$ the longitude of ascending node. The disk mid-plane is chosen to coincide with the $\{x,y\}$-plane.

 \subsection{Physical setting}  

\emph{Unperturbed Stellar cusp}: The initial spherical NSC has a power-law density profile \citep{Merritt2013book}, 
\beq 
\rhoc(r) = \frac{(3-\gamma)\Mbh}{4 \pi \rh^3} \bigg( \frac{\rh}{r}\bigg)^{\gamma} 
\eeq 
and (for $\gamma \neq 2$) the corresponding gravitational potential, 
\beq 
\phic(r) = \frac{G \Mbh}{(2-\gamma)\rh} \bigg( \frac{r}{\rh} \bigg)^{2-\gamma} \,
\eeq 
where $r$ is the radial distance from the central MBH.

Following \citet{Kaur_2018cusp}, we choose a double power-law initial DF $F_0(I,L)$ in 6D phase space, given explicitly as: 
\beq 
\begin{split} 
& F_0(I,L)  = A \, (G \Mbh \rh)^{-3/2} \bigg(\frac{L}{I}\bigg)^{-2 \beta} \bigg( \frac{I}{\sqrt{G \Mbh \rh}} \bigg)^{3 - 2 \gamma} \\[1ex]
& \mbox{with  } A(\beta,\gamma)  = \frac{(3-\gamma)(1-\beta)}{4 \pi^3} \,.    
\end{split} 
\label{F0} 
\eeq 
Here $\beta = 1 -  \sigma_t/(2 \sigma_r)$ is the anisotropy parameter, with $\sigma_t$ and $\sigma_r$ measuring velocity dispersion in tangential and radial directions respectively \citep{Binney_Tremaine_2008}. The normalization constant $A$ is %deduced from the condition of unit cumulative probability within the radius of influence $\rh$ (integrated over full range of $I \in [0,\sqrt{G \Mbh \rh}]$, $L \in [0,I]$, $L_z \in [-L,L]$, and full cycle of angles $\{w,g,h\} \in [0,2\pi]$).  
calculated by requiring that, within the radius of influence $\rh$ (and over the full range of $I \in [0,\sqrt{G \Mbh \rh}]$, $L \in [0,I]$, $L_z \in [-L,L]$, and a full cycle of angles $\{w,g,h\} \in [0,2\pi]$), the integral of $F_0$ is unity.  

%where $A$ is deduced from normalization condition that total number of stars (integrated over full range of $a \in [0,\rh]$, $j \in [0,1]$ and $\ell_z \in [-j,j]$) inside radius $\rh$ is $N_{\rm h}$.  

\emph{Perturbing gas disk}: We consider an evolving gas disk with mass $\Md(t) = \mu(t) \Mbh$ within $\rh$, corresponding to the following density-potential pair (in the language of the more general models constructed in appendix~\ref{app_axis_pot_den}, we use $\gaxis = 3/2$ for the remainder of the paper\footnote{As described further in appendix \ref{app_axis_pot_den}, the specific value of $\gaxis = 3/2$ is also motivated by some theoretical models for AGN gas disks \citep{GilbaumStone22}.}): 
\beq 
\begin{split} 
& \rhod = \frac{ 18 \Md(t)  }{29 \pi \rh^3} \bigg( \frac{\rh}{r} \bigg)^{3/2}  \bigg\{  \delta \bigg(\theta - \frac{\pi}{2} \bigg)   + 
\frac{5}{16} (1 - \left| c_{\theta} \right|)^2
\bigg\}  \\[1ex]
& \phid = \frac{ 36 G \Md(t)  }{29 \rh} \bigg(\frac{r}{\rh} \bigg)^{1/2} \bigg\{ \frac{145}{126} + \left| c_{\theta}  \right| - \frac{5}{42} c_{\theta}^2   \bigg\} .
\end{split}
\label{disk_den_pot}
\eeq 
This corresponds to a shallow profile for disk surface density $\propto R^{-1/2}$, where $R$ is the cylindrical radial coordinate. 
The normalized disk mass $\mu(t)$ evolves over a time $\Tgrow = \TKep(\rh)/\mu_0$, where $\TKep(\rh) = 2 \pi \sqrt{\rh^3/(G \Mbh)}$ represents the dynamical time at $\rh$ and $\mu_0$ is the maximum value attained by $\mu(t)$. For $\mu_0 < 0.1$, $\Tgrow \gtrsim $ a few times $10^{5-6}$ yrs for $\Mbh = 10^{4-8} \Msun$. Later, we will give an explicit (approximate, but astrophysically motivated) model of disk growth in equation~\ref{disk_growth_model} of \S~\ref{sec_CL_calci}. For $\mu_0 < 0.1$, $\Tgrow$ remains at least an order of magnitude longer than both the dynamical and secular timescales within $\rh$, which have an upper limit of $\TKep(\rh)$. Therefore, the disk can be considered to be growing adiabatically.  

\subsection{Secular dynamics}

\begin{figure}
    %\centering
    \includegraphics[width=0.4\textwidth]{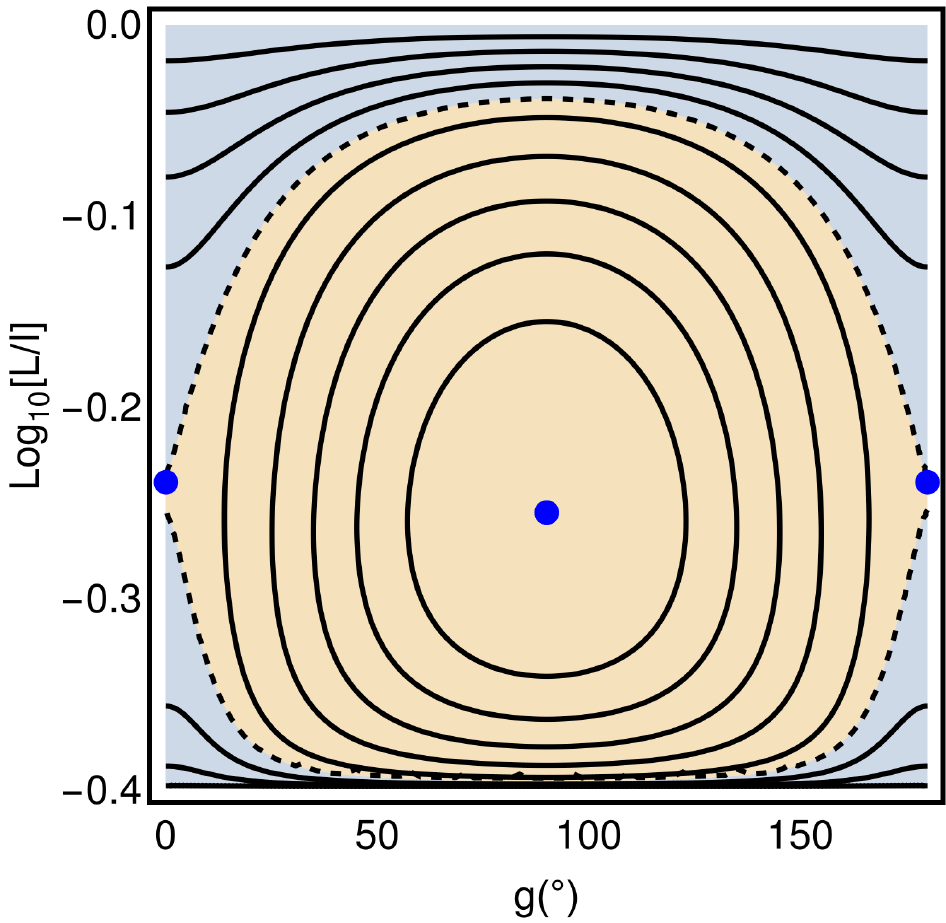}
    \caption{Isocontours of the secular Hamiltonian $H$ in the %$\{g,\ell\}$-plane 
    space of argument of periapsis $g$ and dimensionless angular momentum $\ell$, given a z-component of angular momentum $\ell_z = 0.4$ and disk perturbation strength $\chi = 1.7$ (from $a = \rh$, $\mu = 0.6$, $\gamma = 7/4$). The separatrix orbit, shown as a dashed black contour, divides the plane into three regions -- a libration island (shaded in light orange) containing the librating orbits, and two regions above and below this island corresponding to circulating orbits (shaded in light blue). The blue points correspond to fixed points -- an elliptic point at $g = 90^{\circ}$ and hyperbolic points at $g = 0^{\circ}, 180^{\circ}$.  For this large value of $\ell_z \gg \llc$, stars cannot collisionlessly wander into the loss cone to become TDEs.
   } 
    \label{fig_Hcont_3reg}
\end{figure}

\begin{figure*}
    %\centering
    \includegraphics[width=0.9\textwidth]{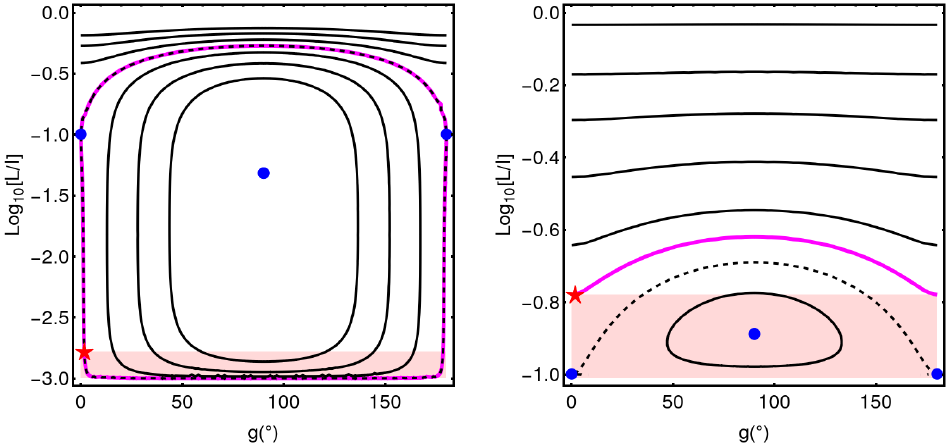}
    \caption{Isocontours of the secular Hamiltonian $H$ (black solid lines) are presented in the same $\{g,\ell\}$-plane as in Fig. \ref{fig_Hcont_3reg}, but now for orbital parameters that permit some collisionless TDEs.  The separatrix orbit is again shown as a dashed black contour, and filled blue circles correspond to fixed points. The TDE loss cone $\ell \leq \llc$ is highlighted as red shaded region. The secular orbit of a collisionless TDE at a given time is shown as a magenta solid line, and its intersection with $\ell = \llc$ represents the fatal Keplerian orbit (red star) for a collisionless TDE. {\it Left panel}: the more typical scenario where a secular TDE orbit is same as the separatrix orbit (for $\lhyp \geq \llc$); the chosen parameters are $\ell_z = 0.001$ and $\chi = 0.28$ (from $a = \rh$). {\it Right panel}: an infrequent scenario where the secular TDE orbit is an upper CO for $\lhyp < \llc$ (as can happen for small $a$); the chosen parameters are $\ell_z = 0.1$ and $\chi = 0.028$ (from $a = 10^{-4}~\rh$). For both cases, we choose $\Mbh = 10^7 \Msun$, $\mu = 0.1$, $\gamma = 7/4$. 
     } 
    \label{fig_Hcont_tde}
\end{figure*}

 The combined gravitational fields of the NSC and the evolving gas disk govern the orbital structure of stars, with the Hamiltonian $\widetilde{H} = v^2/2 - G \Mbh/r  + \phic + \phid$; here $v$ is the magnitude of stellar velocity. For simplicity, we neglect the contribution of the polarization term from deformation of the cusp by the disk. The time-dependent problem we study, with a changing disk mass fraction $\mu(t)$, can be treated as %an effectively time-frozen Hamiltonian
 a sequence of static Hamiltonians in the adiabatic limit. Thus, we orbit-average the Hamiltonian, treating $\mu$ as a constant, to get the secular Hamiltonian, $H = \Phic(I,L) + \Phid(I,L,L_z,g;\mu)$; we drop the contribution from Keplerian energy that does not play any role in secular dynamics. The orbit-averaged cusp and disk potentials are respectively given by $\Phic = \oint \rmd w \; \phic /(2 \pi)$ and $\Phid = \oint \rmd w \; \phid /(2 \pi)$. We describe the details of the transformation to action-angle coordinates from physical phase space, and the resulting calculation of orbit-averaging in appendix~\ref{app_orbit_av_Calci}. The final simplified form of the orbit-averaged cusp potential is 
\begin{subequations} 
\beq
\begin{split}
& \Phic \simeq  \frac{G \Mbh}{(2-\gamma)\rh} \bigg( \frac{a}{\rh} \bigg)^{2-\gamma} (1 + \algam e^2)  ;\\
& \mbox{where} \; \algam = \frac{2^{3-\gamma} \Gamma{(\frac{7}{2}-\gamma)} }{\sqrt{\pi} \Gamma{(4-\gamma)}  } - 1.  \, 
\end{split}
\eeq
This representation of $\Phic$ is approximate, but is correct to within a few percent; see equations~(4.81) and (4.82) of \citet{Merritt2013book}. The approximated form of the orbit-averaged disk potential is likewise  
\beq  
 \Phid \simeq \!  \frac{  G \Mbh \mu(t)  }{ \rh} \! \sqrt{\frac{a}{\rh} } \! \bigg\{
 T_1(e)  + T_2(e,g) \sin^2{i}  + T_3(e,g) \sin{i}     
 \bigg\}
\eeq 
\label{orb_av_phi} 
\end{subequations}
where the functions $T_1(e)$, $T_{2,3}(e,g)$ are given in equation~(\ref{T_val}). In the above expression, we have approximated the disk potential $\Phid$ to an accuracy of $\simeq 4 $ percent; equation~(\ref{Phid_exact}) gives the exact form of $\Phid$. 

In secular dynamics, orbit-averaging the Hamiltonian over mean anomaly $w$ ensures the conservation of $I$. Due to the absence of spherical symmetry, angular momentum $L = I \sqrt{1-e^2}$ is no longer an integral of motion, while axisymmetry allows for the integral $L_z = L \cos{i}$. For a given $\mu$, %the time-frozen approximation considered here,
the orbit-averaged Hamiltonian $H$ is also an integral of motion. This makes the secular dynamics of (nearly Keplerian) axisymmetric systems a fully integrable problem respecting the three integrals of motion $\{I,\, L_z, \, H \}$ \citep{Sridhar_Touma_1999,Vasiliev_Merritt2013}. The orbit-averaged dynamics therefore reduces the problem to an effective 4D phase space defined by normalized action-angle variables $\{\ell = L/I = \sqrt{1-e^2} , \, \ell_z = L_z/I = \ell \cos{i}; \, g , h \}$. As above, the corresponding integrals of motion are $\{H,\ell_z \}$ for a given $\mu$. 

 The secular dynamics of these axisymmetric systems permit two families of orbits: circulating and librating \citep{Sambhus2000,Vasiliev_Merritt2013}, as shown by the isocontours of $H$ in the $\{g,\ell\}$-plane in figure~\ref{fig_Hcont_3reg} for a given $\ell_z$ and disk perturbation strength $\chi$ (defined in equation~\ref{chi_def}). There are four fixed points - elliptic points at $g = \{ 90^{\circ}, 270^{\circ} \}$ corresponding to $\ell = \ell_{\rm ell}$ and hyperbolic points at $g = \{ 0^{\circ}, 180^{\circ} \}$ corresponding to $\ell = \lhyp$.  Circulating orbits (COs; also called tube orbits) are akin to modified rosettes that cover the entire range in apses $g \in [0^{\circ},360^{\circ}]$, coupled with only a moderate change in $\ell$. Librating orbits (LOs; also called saucer orbits) cover a limited range in apses as they librate around the elliptic fixed points at $g = \{ 90^{\circ}, 270^{\circ} \}$, but exhibit huge oscillations in $\ell$. The separatrix orbit (SO), joining the two hyperbolic points, demarcates the boundary of the libration island enclosing all the LOs. It divides the $\{g,\ell\}$-plane into three regions constituting (1) upper COs lying above the libration island, whose minimum attained  $\ell = \lmin > \lhyp$, (2) middle LOs inside the island, and (3) lower COs lying below the island whose maximum attained $\ell=\lmax < \lhyp$.  % give details of this calculation in appendix, along with the normalization of H and introduction of parameter chi. 

 There are two types of timescales associated with secular dynamics of axisymmetric Keplerian star clusters. Firstly, there is the secular timescale $\Tsec$ associated with apsidal precession due to the spherical component of the extended mass distribution, which is mainly contributed by NSC. If $\Mstar(a) = \Mbh (a / \rh)^{3-\gamma}$ is the stellar mass enclosed within $a$, then $\Tsec \sim \Mbh \, \TKep(a)/\Mstar(a)$. The COs that undergo moderate oscillations in $\ell$ apsidally precess over this timescale $\Tsec$. The axisymmetric perturbation from the disk, is quantified by the dimensionless parameter $\chi \sim \Md(a)/\Mstar(a)$, where $\Md(a) = \Md (a/\rh)^{3/2}$ is the disk mass enclosed within $a$. For the exact system we have chosen, the perturbation strength $\chi$ is defined more precisely as (see appendix~\ref{app_actions}):
 \beq 
\chi = \mu \frac{2 - \gamma}{\algam} \bigg( \frac{a}{\rh} \bigg)^{\gamma - 3/2}
\label{chi_def}
 \eeq 
 The disk perturbation introduces the new family of LOs, which complete a cycle of libration in $\ell$ over a slightly longer timescale $\Tlib \sim \Tsec / \sqrt{\chi} = \Mbh \, \TKep(a)/\sqrt{\Mstar(a)\, \Md(a)}$ \citep{Vasiliev_Merritt2013} \footnote{For the sake of completeness, we provide here timescales as functions of dimensionless angular momentum $\ell$ (and also the action $j$, defined later in \S~\ref{sec_CL_calci}), namely $\Tsec(a,\ell) \sim \Mbh \, \TKep(a)/(\Mstar(a)\, \ell)$ and $\Tlib(a,j) \sim \Tsec(a,1)/j$. At the level of rough analytical estimates in the main text at most places, the expression for $\Tlib$ corresponds to $\Tlib(a,\js)$, where $\js$ is the action for the SO and $\js \sim \sqrt{\chi}$ from equation~\ref{jco_approx}. For more rigorous calculations (eg. equation~\ref{Ndot_BB_mu}), we evaluate $\Tlib$ for a given LO numerically. }.  

Our secular Hamiltonian $H$ does not account for relativistic apsidal precession of stellar orbits, that in principle could become important during the part of orbital libration where periapsis is comparable to gravitational radius $r_g$. However, earlier studies (\citealt{Vasiliev_Merritt2013}, their appendix A) indicate that relativistic precession does not significantly alter loss wedge dynamics at large semimajor axis $a \sim \rh$ \footnote{More specifically, equation~(A11) of \citet{Vasiliev_Merritt2013} implies $a/\rh \gtrsim 0.03 (\Mbh/10^7 \Msun)^{4/15}\, (\mu/0.1)^{-1/2}$, which is the necessary condition for relativistic precession to fail to have a leading-order impact on collisionless TDE rates. From the weak power law scalings in this inequality, we conclude that semimajor axes larger than a few per cent of $\rh$ can generally contribute to collisionless TDE rates.}, which are ultimately the semimajor axes that contribute the most to collisionless TDE rates.

\subsection{Collisionless TDEs}
\label{subsec_CL_TDEs}

  Secular orbits with coupled evolution in $\ell$ and $g$%(\ell,g)$
  , as described above, can attain extremely high eccentricities during a part of their circulation or libration cycle. For $\ell_z < \llc(a) = \sqrt{1-\eLC^2} \simeq \sqrt{2 \rtid/a}$, a part of the $\{g,\ell\}$-plane  overlaps with the TDE loss cone (defined by $\ell \leq \llc$). In this case, it is possible for orbits to evolve into the loss cone over a secular time $\Tlib$ 
  even if the initial $\ell \gg \llc$, becoming collisionless TDEs. This extended loss cone, defined as the set of secular orbits that can attain a minimum $\ell = \lmin < \llc$, is the so-called loss wedge \citep{Vasiliev_Merritt2013}, a name that refers to the elongated shape of this region in the $\{\ell,\ell_z\}$-plane \citep{Magorrian_Tremaine_1999}.  Collisionless loss cone (or wedge) repopulation is quite different conceptually from the standard 2B scattering mechanism \citep{FrankRees76}, and we will later investigate their interplay.

As the disk mass fraction $\mu$ grows over a time $\Tgrow$, fresh stellar orbits are captured into loss wedge, and they quickly become TDEs over a time $\Tlib \ll \Tgrow$. Depending upon the parameters $\{\mu,\gamma,a,\ell_z\}$, the secular orbit corresponding to a TDE at a given time can be of two types: (1) a separatrix orbit SO for $\lhyp > \llc$, or (2) an upper CO with $\lmin = \llc$ for $\lhyp < \llc$. We show these two cases in figure~\ref{fig_Hcont_tde}, which also highlights that the region corresponding to lower COs almost vanishes for the low $\ell_z$ values suitable for TDE formation. The intersection of the $H$ contour of the relevant secular orbit with $\ell = \llc$ gives the Keplerian orbit that results in a TDE \footnote{Note that the actual Keplerian orbit of a collisionless TDE can be affected by two additional effects not accounted for here: (1) relativistic %apsidal 
precession, which becomes most important for low $\ell$ orbits near loss cone, and (2) discrete jumps in $\{\ell,g\}$ during the libration cycle (i.e. going beyond the secular approximation of the current study). Though these two effects can influence the resulting TDE orbits, we expect their influence on the formation rate of collisionless TDEs to be sub-dominant \citep{Vasiliev_Merritt2013}.} (for a given combination of $\{\mu,\gamma,a,\ell_z\}$). As most collisionless TDEs arise from larger $a$ (see equation~\ref{Ntde} in section~\ref{sec_CL_calci}), the condition $\lhyp > \llc$ is satisfied much more frequently and the SO generally represents the secular orbit for a collisionless TDE. Note that this description assumes that stellar orbits can perform secular librations, remaining uninterrupted by relaxation of orbital elements due to 2B scatterings. %We call these resulting TDEs as collisionless TDEs. 
This regime of dynamics and the conditions of its validity are discussed later in this section.%next. 

\emph{Orbital Characteristics of Collisionless TDEs}: In the limit of secular dynamics, collisionless TDEs tend to attain small periapsidal angle\footnote{Here we describe the dynamics within the apsidal range $g \in [0^{\circ},180^{\circ})$, because the nature of secular motion is %exactly translated 
identical in the remaining range $g \in [180^{\circ},360^{\circ})$. Hence, $\gtde$ should be interpreted as modulo $180^{\circ}$. } $\gtde \approx 0^{\circ}$ (corresponding to the red star in figure~\ref{fig_Hcont_tde}), so that the line of apses roughly aligns with the disk plane. This is due to the peculiar geometry of the $H$-contours of the SO (left panel in figure~\ref{fig_Hcont_tde}) especially near the lower boundary, which implies a steep decline in $\gtde$ as $\llc$ increases with respect to $|\lz|$. Note the directional preference for a small, but positive $\gtde$ (because of the retrograde precession of apsides of COs above the separatrix), which implies the disruption of star (at periapsis) only occurs just after crossing the disk at the inner node. This geometrical preference may have important consequences for the observational appearance of collisionless TDEs in AGN disks, which we discuss below.

\begin{figure}
 \includegraphics[width=0.55\textwidth]{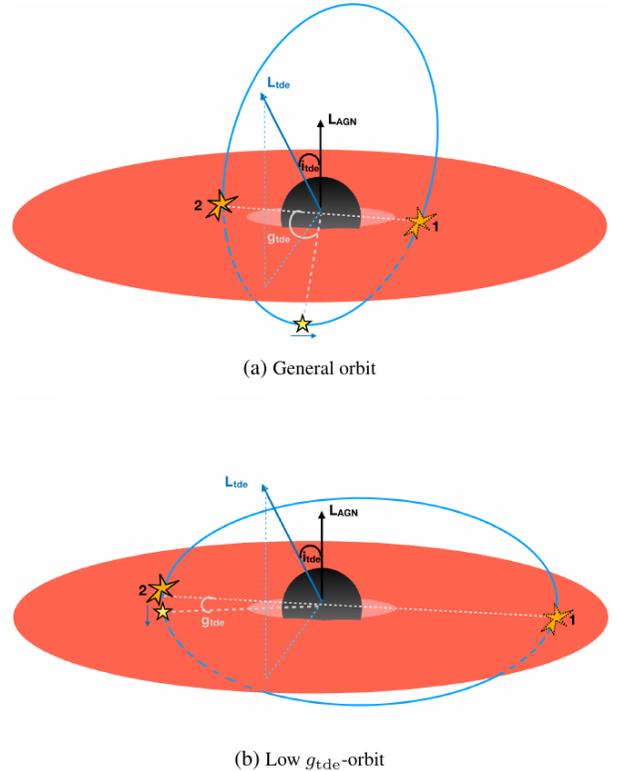}
    \caption{Two example orbital geometries for the most bound orbit (in blue) of debris streams resulting from a collisionless TDE, occurring in an AGN disk (shown in red) around a central MBH (in black). After the stellar disruption occurs at periapsis (shown as a yellow star), the resulting stream interacts with the disk at the two impact points (or nodes), shown as orange impact flashes and labelled as 1 and 2. Panel (a) presents an orbit with a general orientation ($\gtde$) of lines of apsides and nodes.  Panel (b) shows the case of small periapsidal angle $\gtde$, which ensures that the impact point occurs at a larger distance near stream apocenter. In the limit of secular, non-relativistic dynamics, almost all collisionless TDEs result in the latter scenario, with implications for the resulting transient \citep{Chan_Piran2021}. The orbital eccentricity in this sketch is far smaller than its actual value, and is adopted for a clear presentation.   
    }
    \label{fig_sketch}
\end{figure}

As small and positive $\gtde$ values are favored, the resulting debris streams will generally first encounter the disk in a near-apocentric impact (as shown in panel (b) of figure~\ref{fig_sketch}). These more distant apocentric impacts may feature stream momentum currents comparable to those of the pre-existing AGN disk, potentially dissipating stream kinetic energy in shocks at this first impact, or alternatively enhancing the spread of stream trajectories. In the latter case, the lower density stream would likely slam to a halt during its next nodal passage near pericenter.  In either case, Compton cooling of the post-shock material may produce a hard X-ray/soft $\gamma$-ray flare \citep{Chan_Piran2021}. Hence, collisionless TDEs in AGN can have distinct high-energy signatures, that may help to identify them over underlying AGN disk emission. We expect this picture, based on secular non-relativistic dynamics, to remain valid for low-mass MBHs for which relativistic apsidal precession remains small. An additional caveat is that discrete jumps in apsides during a libration cycle near $\rh$, where secular dynamics is only marginally valid, can lead to collisionless TDEs with a relatively large $\gtde$. However, while these caveats complicate the calculation of a $\gtde$ distribution, we do not expect them to significantly influence the occurrence rates of collisionless TDEs.

\emph{Intervention by 2B scatterings}: Stellar orbits generally undergo relaxation in energy and angular momentum over the two-body relaxation time $\Tbb(a) \sim \TKep(a) (\Mbh/\mstar)^2/(N_{\star}(a) \lnL )$, where $N_{\star}(a)$ is the number of stars with semi-major axis $\leq a$ and $\lnL \sim \ln{(\Mbh/\mstar)}$ is the Coulomb logarithm \citep{Binney_Tremaine_2008}. However, the relaxation for small $\ell$ and/or $\ell_z$ occurs over much shorter timescales, $\Tl(a,\ell) \sim \ell^2 \, \Tbb(a)$ and $\Tlz(a,\ell_z) \sim \ell_z^2 \, \Tbb(a)$ respectively \citep{Merritt2013book,Vasiliev_Merritt2013}. So, the $\ell_z$-relaxation timescale at the loss wedge boundary $\ell_z = \llc(a)$ is $\Tlc(a) \sim \llc^2 \, \Tbb $, which is similar to the $\ell$-relaxation timescale at the loss cone boundary $\ell = \llc(a)$ \footnote{Of course, the concept of relaxation in $\ell$ is only an approximate notion for an axisymmetric system, as $\ell$ is no longer an integral of motion (see \citealt{Vasiliev_Merritt2013} for an exact treatment).}. The secular channel of supplying collisionless TDEs, discussed above, can remain functional if secular orbits inside the loss-wedge can librate freely without interruptions by 2B scatterings. This is possible if the loss-wedge relaxation time $\Tlc(a)$ is much longer than the libration time $\Tlib$. Since most collisionless TDEs are sourced by large $a \sim \rh$, it is sufficient to compare these timescales at $a=\rh$. As the secular timescale $\Tlib$ becomes comparable to the dynamical timescale $\TKep$ at $a \sim \rh$, the condition for the simple validity of the above collisionless channel becomes $\Tlc(\rh) > \TKep(\rh)$ \footnote{This condition is equivalent to an empty loss-cone regime, usually discussed in context of spherical star clusters.}. This can be written more explicitly as $\Mbh \, \rtid/(\mstar \, \rh) > \lnL/2$, which further reduces to $\Mbh \gtrsim 2 \times 10^6 \Msun$ for $\lnL = 10$ and $\rh = 2~{\rm pc} \, (\Mbh/(4 \times 10^6 \Msun))^{3/5}$, corresponding to $\Mbh-\sigma$ relation \footnote{This captures the observed properties of MBHs and galactic nuclei on average, because the exponent of $\sigma$ generally lies in the range $\sim 4 -6$ \citep{Kormendy_Ho_2013,McConnell2013}. } of the form $\Mbh \propto \sigma^5$. Hence, the collisionless TDEs will contribute most significantly only for relatively large central MBHs.              

\section{Collisionless TDE rates channelled by growing gas disk}
\label{sec_CL_calci}

In this section, we first evaluate the non-linear response of the cusp DF to an adiabatically evolving gas disk. We then utilize this deformed DF to evaluate the time-dependent flux of collisionless TDEs.   

We assume a simple evolutionary model for the disk mass, such that $\mu$ grows exponentially over a time $\Tgrow$ for $t < 0$, reaching a maximum mass ratio $\mu_0$ at $t = 0$. After reaching its peak mass, the disk enters a short-term decay phase lasting for a time $\Tdecay$ after which $\mu=\mu_r$ retains a constant value: 
\beq
\mu(t) = \begin{cases}
    & \mu_0 \exp{(t/\Tgrow)} \; , \; t \leq 0 \\
    & (\mu_0 - \mu_r) \exp{(-t/\Tdecay)} + \mu_r \; , \; t > 0    .
\end{cases}
\label{disk_growth_model}
\eeq 
Here we take $\Tgrow = \TKep(\rh)/\mu_0$, so that more massive disks assemble faster.  We speculate that this correlation could arise from (i) larger progenitor molecular clouds exerting higher non-axisymmetric torques (at $r > \rh$), forcing higher rates of infall inside $\rh$, and (ii) non-axisymmetric instabilities arising in particularly massive gas disks, leading to global intra-disk torques \citep{Thompson+05}. The influence of the progenitor cloud's mass on disk assembly timescales has not been systematically investigated in literature, though the above form of $\Tgrow$ gives disk-growth timescales $\sim $ a few $0.1 - 1$~Myr (for $\mu_0 \simeq 0.1$), agreeing with simulations at an order of magnitude level \citep{Alig_2011,Mapelli_2012,Trani_2018,Generozov_Nayakshin_2022}. Also, there are other physical reasons that motivate for an inverse dependence on disk mass (or cloud's mass). Firstly, the induced secular precession can assist self-intersection of streams, leading to faster energy dissipation by shocks. \citet{Trani_2018} highlight the significance of secular precession for disk formation, though the NSC is the main driver of precession there. Further, they found that bigger clouds also provide a greater initial shear among the debris streams, that further assists their intersection. While we caution the reader about underlying uncertainties in the growth timescale due to a multitude of factors (e.g. the unconstrained properties of central molecular clouds in general galactic nuclei, the dependence of disk formation on initial properties and orbit of cloud, and the possibility of more complicated scenarios leading to disk formation, like collisions of two clouds; \citealt{Hobbs_2009,Mapelli_Trani_2016,Trani_2018,Tartenas_2020}), we will see that the choice of $\Tgrow$ does not affect the total number of collisionless TDEs.  The total TDE count we predict is therefore robust to these uncertainties, but the resulting TDE rate {\it does} have an inverse proportionality on $\Tgrow$.      

The decay of the disk mass happens over a time $\Tdecay$. As we will see below, collisionless TDEs can only occur during the growth phase, and the form of decay time can not influence their rate\footnote{Later, we choose a decay time corresponding to dynamical time at $\rh$; that is $\Tdecay = \TKep(\rh)$ while computing collisional TDE rates in \S~\ref{subsec_collisional_rates}.}. These later phases, after growth ceases, might correspond to mass-loss due to star formation in, or outflows from, the disk.

The creation of a libration island in the $\{g,\ell\}$ plane is a hallmark of disk-induced asphericity. During the disk growth phase, this island expands, with COs being captured and transformed into LOs that can eventually fuel collisionless TDEs. These purely collisionless TDEs emerge only during the growth phase, and are not possible afterwards, as the island subsides and shrinks during the decay of disk. % and LOs are released back into circulating phase space.
After the growth phase ends at $t=0$, only collisional evolution of stellar orbits (triggered by 2B scatterings) refill the loss wedge, producing TDEs over the longer relaxation time $\Tbb$. As we are primarily interested in evaluating collisionless TDE rates, we solve for the deformed DF only during the growth phase $t \leq 0$. 

It is imperative to track the adiabatic evolution of the cusp DF $F$ in suitable variables, namely the instantaneous action-angles $\{j,\phi_j\}$ of the system rather than $\{\ell,g\}$. The deformed system is expected to quickly phase-mix over a secular time $\Tlib$, so that $F$ does not depend on angle $\phi_j$ \citep{Sridhar_1996}. We define the actions $j$ for orbits in the three regions of $\{g,\ell\}$-plane (figure~\ref{fig_Hcont_3reg}) in a standard way, which we sketch here. The upper COs are denoted by an action $\jc$ (above the upper boundary of SO labelled with action $\js$), middle LOs with an action $\jl$ lying inside the island bounded by SO, and lower COs with an action $\jc'$ below the lower boundary of the SO\footnote{For the parameters of interest with low $\ell_z \leq \llc$, the region of lower circulating orbits $\jc'$ occupies a negligible volume in phase space and hence is not important.}. At a given $\mu$, the upper and lower COs interacting with the SO correspond to actions $\jco$ and $\jco'$, and the action for SO is $\js = \jco - \jco'$.  As usual, the value of the action for a CO equals the area under the corresponding $H$-contour in phase space; more explicitly, $\jc = (180^{\circ})^{-1} \int_{0^{\circ}}^{180^{\circ}} \ell(H; \mu, a, \ell_z) \, \rmd g $. The action for a LO equals the area bounded within the corresponding closed $H$-contour, i.e. $\jl = (180^{\circ})^{-1} \oint \ell(H; \mu, a, \ell_z) \, \rmd g $. We more fully describe how to evaluate the actions in appendix~\ref{app_actions}.

\subsection{Adiabatically deformed DF} 
\label{sec_deformed_DF_Calci}

We utilize adiabatic conservation of actions to evaluate non-linearly deformed DF $F(a,j; \mu)$\footnote{The non-dependence of the deformed DF on the action $\ell_z$ results from adiabatic conservation of actions because the initial DF $F_0$ represents a non-rotating system, independent of $\ell_z$.}. Since we are only concerned about the growth phase of the disk, we can refer to time via its monotonic function $\mu$. It is interesting to note that any slowly growing perturbation is not really adiabatic near the separatrix due to the slowly 
varying phase $g$, so actions of orbits crossing %interacting with
the SO do in fact change as they are captured into the libration island or released into the lower CO region. However, the detailed analysis of motion near the SO by \citet{Henrard1982} revealed the underlying simplicity of this process, identifying the effective invariance of actions once that one allows for change in their geometric definition upon interaction with SO. \citet{Sridhar_1996} (hereafter ST96) provide a useful description of this process in terms of DFs, rather than probabilities of capture and escape. We follow their approach to determine the deformed DF $F(a,j; \mu)$ at any given $\mu$.   

During the growth phase (i.e. increasing $\mu$) of interest to us, both the upper and lower boundaries of the SO move upwards; however, the upper boundary moves faster so that the island expands (and $\js$ increases) with increasing $\mu$ \footnote{We verified this velocity hierarchy by checking the rates of change ($\vplus$ and $\vminus$) of actions corresponding to the upper and lower boundaries of SO, with different disk perturbations $\mu$. This hierarchy also seems intuitively correct. The upper boundary of the SO moves upwards with $\mu$ because high inclination orbits are captured into the libration island as they are strongly impacted by the disk. Further, the lower boundary of the SO, near low-inclination orbits, also moves upwards as the orbits very closely aligned with disk undergo only circulation, similar to rosette orbits in spherical geometry. }.  We define \emph{velocities} of the upper and lower boundaries of SOs: $\vplus = \p \jco/\p \mu$ and $\vminus = \p \jco'/ \p \mu$. As our problem features an expanding island with $\vplus > \vminus > 0$, it corresponds to case (b) of Table 1 in ST96. Consequently, a CO (in the upper region) interacting with the SO has a probability $\vminus/\vplus$ to escape into the lower CO region, and a probability $(1 - \vminus/\vplus)$ to get captured into the libration island \citep{Henrard1982}. In the language of ST96, the DF $F(a,\jco;\mu)$ of the upper interacting CO (belonging to the shrinking region) is the independent quantity, that equals and hence determines the DFs for both SO and lower interacting CO. More exactly, $F(a,\js;\mu) = F(a,\jco;\mu)$ and $F(a,\jco',\mu) = F(a,\jco;\mu)$. The numerical values of the action and the corresponding DF of orbits away from the SO, in the rest of the phase space, remain the same at a time $\mu$. Based on this basic picture, we outline below the steps to determine the deformed DF in all of action space at any $\mu$ (during $t \leq 0$), given knowledge of the initial DF $F_0(a,\ell)$ \footnote{Note that earlier in \S~2, we denoted the initial DF $F_0$ as the function of $(I,L)$, so strictly speaking it should be denoted with a different symbol here, but we slightly abuse the notation for simplicity.}.  %Here we utilize the prescription of capture in terms of DF formulated by \citet{Sridhar_1996}.    

\begin{itemize}
    
\item \emph{Upper circulating orbits}: Applying conservation of action is trivial for those upper COs that have not yet interacted with the SO. At a given $\mu$, the action $\jc$ of a circulating orbit just represents the initial value of $\ell$ associated to that orbit. Hence, $F(a,\jc;\mu) = F_0(a,\jc)$. %This procedure can be applied to COs in the lower region of phase space as well.    

\item \emph{Separatrix}: For the separatrix orbit with $\js(a,\ell_z;\mu) = \jco-\jco'$, the effective conservation of actions (from ST96) implies $F(a,\js;\mu) = F(a,\jco;\mu) = F_0(a,\jco)$. The latter equality just utilizes the result for upper COs.   

\item \emph{Librating orbits}: Once an upper CO is captured inside the libration island at time $\muc$, its action value $\js(\muc)$ remains fixed by trivial application of action conservation. Geometrically, the orbit submerges deep inside the expanding island as $\mu$ increases. At a given $\mu$ and for a given LO with action $\jl$, we solve for the $\muc$ at which it was captured by separatrix by numerically solving the equation $\js(\muc) = \jl$. Then we evaluate the action value $\jco(\muc)$ of the corresponding initial CO, that was captured at $\muc$. Hence the corresponding value of DF for LO $F(a,\jl;\mu) = F_0(a,\jco(\muc))$.   

\item \emph{Lower circulating orbits}: Similarly, for a given lower CO orbit with action $\jc'$ at $\mu$, we can determine the moment of escape $\mue$ of the initial upper CO by numerically solving $\jco'(\mue) = \jc'$. Again, the procedure of ST96 implies that the DF $F(a,\jc';\mu)  = F_0(a,\jco(\mue))$.  

\end{itemize}

The above procedure fully determines the adiabatically deformed DF $F(a,j;\mu)$ at all times. We then utilize it to evaluate the rates of collisionless TDEs below.  

\subsection{Collisionless TDE rates}
\label{sec_CL_TDE_rates}

\begin{figure*}
    \centering
    \includegraphics[width=1\textwidth]{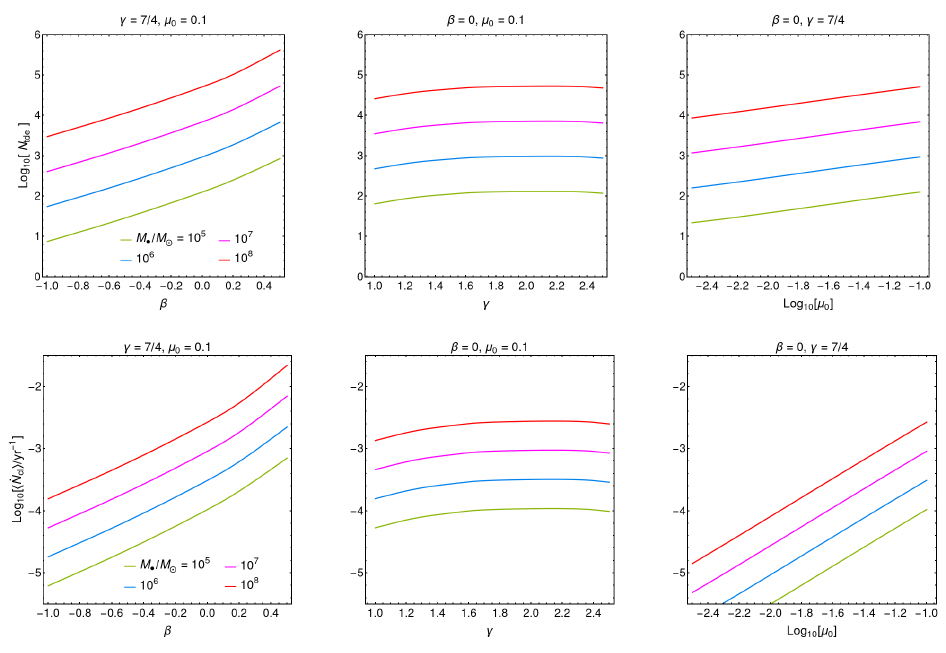}
    \caption{
    Total number of collisionless TDEs $\Ntde$ (upper panel) and average rates of collisionless TDEs $\NdotCLav$ (lower panel) as functions of $\{ \Mbh, \mu_0, \beta , \gamma \}$. Both $\Ntde$ and $\NdotCLav$ are always increasing functions of MBH mass $\Mbh$ (shown in color-coding). The first column shows variation with respect to the NSC anisotropy parameter $\beta$ (with fixed $\mu_0 = 0.1$, $\gamma = 7/4$); the second column with the stellar density slope $\gamma$ (fixed $\beta = 0$, $\mu_0 = 0.1$); the third column with the disk mass ratio $\mu_0$ (fixed $\beta =0$, $\gamma = 7/4$). While $\Ntde$ and $\NdotCLav$ are monotonically increasing with $\{\Mbh, \mu_0, \beta \}$, their dependence on $\gamma$ is non-monotonic, and not very sensitive. The disk growth times $\Tgrow = \{ 0.3, 0.76 , 1.9 , 4.8   \}$ Myr for $\log_{10}[\Mbh/\Msun] = \{5,6,7,8\}$.  The highest collisionless TDE rates, and the largest enhancements over standard TDE rates from two-body scatterings, are both obtained for large $M_\bullet$ and $\mu_0$.
    } 
    \label{fig_NtdeCL_av}
\end{figure*}

The fraction of phase space drained by TDEs before a given $\mu$ is reached depends on the \emph{critical} orbit, with maximum action $\jcrit(a,\ell_z;\mu)$, whose minimum angular momentum $\lmin \leq \llc$. All the orbits with lower action values $j < \jcrit$ would have already fed their stars into the TDE loss cone in the past, so $j = \jcrit$ represents the secular orbit of collisionless TDEs at a given $\mu$. Since $\lhyp > \llc$ for most of the parameter space $\{a, \ell_z, \mu \}$, this orbit corresponds to the SO with $\jcrit = \js$ (see figure~\ref{fig_Hcont_tde} and discussion in \S~\ref{subsec_CL_TDEs}). At smaller $a$ where the loss cone becomes relatively large, it is possible to have $\lhyp < \llc$, and the critical orbit will then correspond to an upper CO orbit whose $\lmin = \llc$; only a small minority of TDEs come from such small $a$, however. Hence the cumulative number of collisionless TDEs supplied uptil $\mu$: 
\beq 
\begin{split}
& \Ntde(\mu; \Mbh, \gamma, \beta) = \\
& \quad 4 \pi^3 (G \Mbh)^{3/2} N_{\rm h} \int_{0}^{\rh} \rmd a \, \sqrt{a} \int_{-\llc}^{\llc} \rmd \ell_z \int \rmd j    F(a,j,\mu)
\end{split}
\label{Ntde}
\eeq 
where $N_{\rm h} = N_{\star}(\rh) = \Mbh/\mstar$ is the number of stars with $a \leq \rh$. Here integration over $j$ includes the entire action space, with $j \leq \jcrit(a,\ell_z;\mu)$. We evaluate the total number $\Ntde(\mu_0)$ of collisionless TDEs channelled by a growing disk with maximum mass $\mu_0$ and investigate its dependence on various parameters in figure~\ref{fig_NtdeCL_av} (upper panels). Figure \ref{fig_NtdeCL_av} (lower panels) also shows the average rate of TDEs $\NdotCLav = \Ntde(\mu_0)/(4 \Tgrow)$ implied\footnote{Here we choose the time-interval $4 \Tgrow$ to get the average TDE rate $\NdotCLav$, because most of the TDEs ($\sim 80\%$) occur during this time in our chosen disk evolution model (equation~\ref{disk_growth_model}) for an isotropic cluster ($\beta = 0$). We then use this as a fiducial averaging timescale for all systems with general $\beta$.} by our chosen model for the disk growth timescale $\Tgrow = \TKep(\rh)/\mu_0$. Our analysis finds that $\Ntde$ and $\NdotCLav$ are quite sensitive to the parameters $\{ \Mbh,  \mu_0 , \beta \}$, while their dependence on $\gamma$ is rather weak. Throughout this work, we consider Solar-type stars with mass $\mstar = \Msun$, radius $\Rstar = \Rsun$, and a nuclear star cluster with radius of influence $\rh = 2\, {\rm pc} \, ( \Mbh/(4 \times 10^6 \Msun) )^{3/5} $ (as mentioned in \S~\ref{subsec_CL_TDEs}). We describe in detail the trends of $\Ntde(\mu_0)$ and $\NdotCLav$ with respect to various parameters of interest below (and derive analytic scalings that are quoted in the following enumerated summary later in this sub-section).  We caution that while our estimates for $\NdotCLav$ ultimately depend significantly on our parametrization of $T_{\rm grow}$, our estimates for $\Ntde$ are ultimately independent of the detailed time evolution of the AGN disk, provided it is adiabatic.

\begin{itemize}
    \item[(1)] \emph{MBH mass}: Both the number and the rates of collisionless TDEs increase for MBHs with higher mass $\Mbh$, with $\Ntde \propto \Mbh^{13/15}$ and $\NdotCLav \propto \Mbh^{7/15}$. This mainly arises due to the larger number of stars $N_{\rm h} \propto \Mbh$ at the radius of influence $\rh$, though the size of the loss cone $\llc(\rh) \propto \sqrt{\rtid/\rh} \propto \Mbh^{-2/15}$ decreases slightly for higher $\Mbh$. The weaker dependence of $\NdotCLav$ on $\Mbh$ arises due to longer disk growth timescales $\Tgrow(\rh) \propto \TKep(\rh) \propto \Mbh^{2/5}$ for higher $\Mbh$.  Because the largest MBHs produce the highest AGN TDE rates, a general relativistic treatment of the disruption criterion \citep{Kesden12} should ultimately be applied to study AGN TDE rates near the Hills mass \citep{Hills_1975} of $M_\bullet \sim 10^8 M_\odot$, but this is beyond the scope of the present work.

    \item[(2)] \emph{Disk mass}: A higher disk mass ratio $\mu_0$ leads to a larger libration island ($\js \propto \sqrt{\mu_0}$; see equation~\ref{jco_approx}) and hence more trapped orbits, leading to a higher total number and average rate of collisionless TDEs. For an initially isotropic cluster ($\beta =0$), our numerical results give $\Ntde \propto {\mu_0}^{1/2}$ and $\NdotCLav \propto \mu_0^{3/2}$. The stronger dependence of $\NdotCLav$ on $\mu_0$ is due to the particular model of disk growth timescale $\Tgrow \propto \mu_0^{-1}$. Later, we will see from analytical  approximations that the dependence on $\mu_0$ is convolved with the anisotropy parameter $\beta$, so that $\Ntde \propto {\mu_0}^{(1 - 2 \beta)/2}$ and $\NdotCLav \propto \mu_0^{(3 - 2 \beta)/2}$ (equations~\ref{Ntde_approx2} and \ref{NdotCLav_approx}).

    \item[(3)] \emph{Anisotropy parameter}: Both the number and the rate of collisionless TDEs are strongly increasing functions of $\beta$, varying roughly by two orders of magnitude for $\beta \in [-1,0.5]$. This is because a higher radial anisotropy ensures a higher number of trapped orbits within the libration island. Equations~\ref{Ntde_approx1} and \ref{NdotCLav_approx} highlight the analytical form of dependence on $\beta$, which is convolved with $\mu_0$ and $\gamma$.  

\item[(4)] \emph{Density index $\gamma$}: Both $\Ntde$ and $\NdotCLav$ have a non-monotonic and weak dependence on $\gamma$. Both these quantities generally increase with increasing $\gamma$, as the perturbative effect of the disk at $\rh$ increases for a steeper cluster density profile, leading to a larger width of the libration island (higher $\js$ for higher $\gamma$ at $a = \rh$ in equation~\ref{jco_approx}). For extreme density profile slopes ($\gamma \simeq 2.5$), $\Ntde$ and $\NdotCLav$ slightly decrease due to the reduced number of stars near $\rh$, which contribute dominantly to TDE fluxes \footnote{The analytical integration leading to equation~\ref{Ntde_approx1} suggests that the total number of TDEs contributed by semi-major axes near $a$ is proportional to $a^{(\beta + \frac{1}{2}) (\frac{3}{2}-\gamma)+1}$. Hence, higher $a$ contribute higher number of TDEs for $\gamma \in [1,2.5]$ and $\beta \in [-1,0.5]$. }.

\end{itemize}

\noindent
\emph{Analytical estimates}: Here we derive an approximate analytical expression for $\Ntde(\mu)$ to understand better the above dependence on various parameters. In the below calculation, we only consider the more prevalent scenario where the SO represents the secular orbit of a collisionless TDE at time $\mu$, such that $\jcrit = \js$. The action of the SO can be approximated as follows, up to a high precision of $\sim 10^{-3}$: 
\beq 
\js \simeq 0.585 \sqrt{\chi} \quad, \quad \chi = \mu \frac{2 - \gamma}{\algam} \bigg( \frac{a}{\rh} \bigg)^{\gamma - 3/2},
\label{jco_approx}
\eeq 
for $|\ell_z | \leq 0.01$ (relevant for the loss wedge) and $\chi \in [10^{-4}, 1]$, (which covers the physically interesting range of parameters $a/\rh \in [0.01,1]$, $\gamma \in [1.2,2.5]$, $\mu \in [0.01,0.1]$).

For the small $\ell_z$ values lying in the loss wedge, we can neglect the lower COs (see figure~\ref{fig_Hcont_tde}), so that we can estimate actions and the deformed DF $F$ approximately using $\jco' = 0$ in the procedure outlined in \S~\ref{sec_deformed_DF_Calci}. This leads to a trivial identity mapping the current action $j$ (for both LOs and COs) to the initial $\ell$, such that $F(a,j;\mu) = F_0(a,j)$. We have the initial DF $F_0$ from equation~\ref{F0}, giving:
\beq 
F_0(a,j) = A(\beta,\gamma) \, (G \Mbh \rh)^{-3/2} j^{-2 \beta} \bigg( \frac{a}{\rh} \bigg)^{\frac{3}{2} - \gamma},
\label{F0_simple} 
\eeq 
where $A(\beta,\gamma)$ is the normalization constant.  %condition that total number of stars (integrated over full range of $a \in [0,\rh]$, $j \in [0,1]$ and $\ell_z \in [-j,j]$) inside radius $\rh$ is $N_{\rm h}$.  

Using the above approximations, the integral of equation~\ref{Ntde} can be solved straightforwardly:
\begin{equation*}
\Ntde \simeq N_{\rm h} \sqrt{\frac{\rtid}{\rh}} \mu^{\frac{1 - 2 \beta}{2}} \mathF(\beta,\gamma)
\end{equation*}
with the factor $\mathF(\beta,\gamma)$ given as: 
\beq 
\begin{split}
& \mathF(\beta,\gamma)  = 0.46 \times 2^{3/2}\\[1ex]
& \quad  \frac{(3 -\gamma)(1-\beta)}{1 + (\frac{3}{2}-\gamma)(\frac{1}{2} +\beta )  }  \frac{(0.585)^{1-2 \beta}}{1- 2 \beta} \left( \frac{ 2 - \gamma}{\algam}  \right)^{\frac{1 - 2 \beta}{2}}  
\end{split}
\label{Ntde_approx1}
\eeq 
where $\beta < 1/2$. This analytical expression reproduces numerical results extremely well, with the fractional difference of at most $10$ per cent in the range $\beta \in [-1,0.3]$, $\gamma \in [1.2,2.5]$ for all $\{\mu,\Mbh\}$ values of interest. Note that we have multiplied our actual analytical result by a numerical prefactor of 0.46 to better match with the numerical evaluation of the integral of equation~\ref{Ntde}.

To deduce a rough analytical dependence of $\mathF$ on $\beta$, we choose $\gamma= 7/4$, corresponding to a Bahcall-Wolf density profile (\citealt{Bahcall_Wolf_1976}) as a fiducial reference (though we recall the relatively weak sensitivity of $\Ntde$ on $\gamma$) and define $\mathF_0(\beta) = \mathF(\beta,7/4)$, which can simply be expressed as:
\beq 
\mathF_0 \simeq \frac{6.5 (1-\beta)}{(1- 2 \beta)(7/2 - \beta)},
\eeq 
with precision upto $4$ per cent. Note that the variation of $\mathF(\beta,\gamma)$ with $\gamma$ is $\mathcal{O}(1)$, with $\mathF/\mathF_0 \in [0.75,2]$ for $\beta \in[-1,0.4]$ and $\gamma \in [1.2,2.4]$. For an isotropic system, we have $\mathF_0(0) = 1.8$.     

In order to check the explicit dependence of $\Ntde$ on parameters of interest, we can cast equation~\ref{Ntde_approx1} into the following form: 
\beq 
\Ntde \simeq 1.2 \times 10^4 M_7^{13/15} \mu^{\frac{1 - 2 \beta}{2}} \mathF(\beta,\gamma),
\label{Ntde_approx2}
\eeq 
where we have taken stellar parameters for a Solar-type star, and assumed $\rh \propto \Mbh^{3/5}$. For an isotropic cluster with $\beta = 0$, we have $\Ntde \simeq 6.8 \times 10^3 M_7^{13/15} (\mu/0.1)^{1/2}$ using $\mathF \simeq  \mathF_0(0)$.   % (i.e. stellar mass $\mstar = \Msun$, radius $\Rstar = \Rsun$). 

Using equations~\ref{Ntde_approx1} and \ref{Ntde_approx2}, the average rate $\NdotCLav = \Ntde(\mu_0)/ (4 \Tgrow)$ of collisionless TDEs is: 
\beq 
\begin{split} 
\NdotCLav &= \frac{N_{\rm h}}{4 \TKep(\rh)} \sqrt{\frac{\rtid}{\rh}} \mu_0^{\frac{3 - 2 \beta}{2}} \mathF(\beta,\gamma) \\[1ex]
  &= 1.5 \times 10^{-2} {\rm yr}^{-1} M_7^{7/15} \mu_0^{\frac{3 - 2 \beta}{2}} \mathF(\beta,\gamma).
\end{split} 
\label{NdotCLav_approx}
\eeq 
For an isotropic cluster, we have $\NdotCLav = 8.5 \times 10^{-4} {\rm yr}^{-1} M_7^{7/15} (\mu_0/0.1)^{3/2}$. 

\medskip

\begin{figure}
    \centering
    \includegraphics[width=0.45\textwidth]{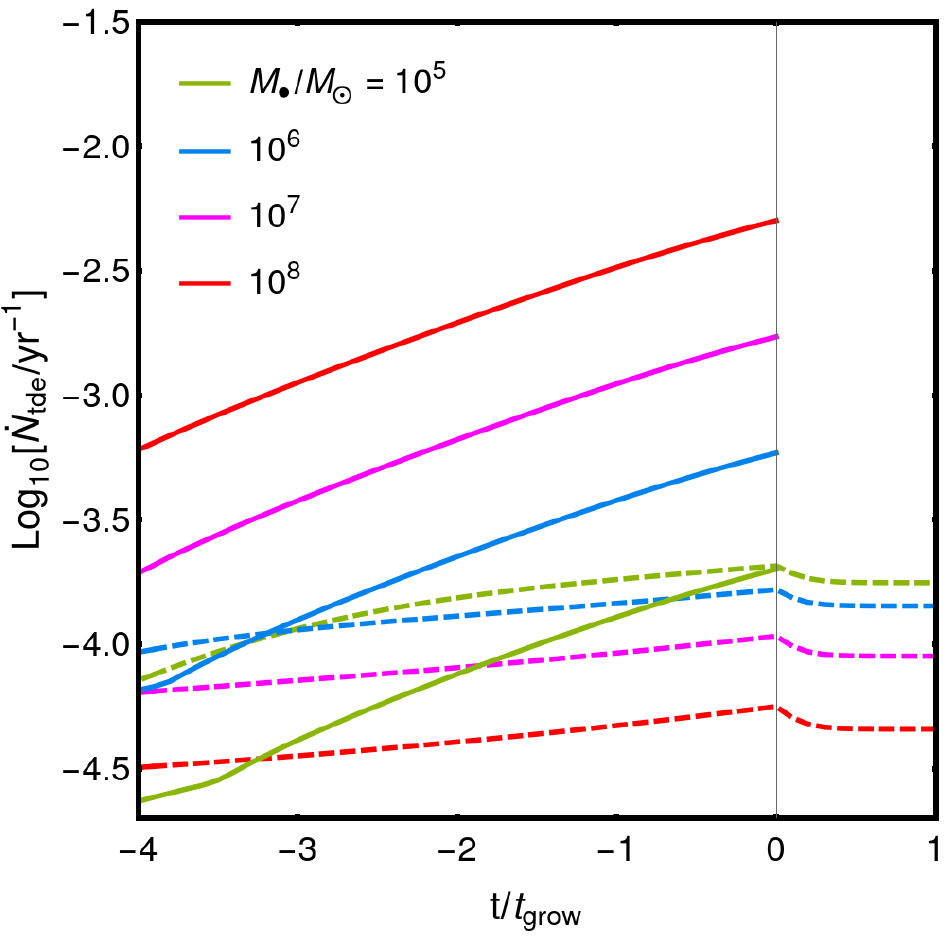}
    \caption{
    Time evolution of TDE rates for a galactic nucleus with central MBH of mass $\Mbh/\Msun = 10^{5-8}$ (shown in different colors) and a surrounding isotropic NSC ($\beta =0$) with density profile power-law index $\gamma = 7/4$. Collisionless TDE rates (solid lines) are higher for larger $\Mbh$ and increase with time as the disk mass $\mu$ grows, until cutting off at $t = 0$. The collisionless contribution to the TDE rate vanishes as soon as the disk begins to lose mass ($t > 0$). Here we consider a maximum disk mass ratio $\mu_0 = 0.1$. Collisional TDE rates (dashed lines) increase only moderately with increasing $\mu$ due to the expanding loss wedge (see \S~\ref{subsec_collisional_rates}). During the decay phase ($t>0$), only collisional TDEs occur. The collisionless enhancement is most notable for high mass MBHs with $\Mbh \gtrsim 10^6\Msun$, while collisional TDEs are expected to dominate for lower mass MBHs. Here the disk growth times $t_{\rm grow} = \{ 0.3, 0.76 , 1.9 , 4.8   \}$ Myr for $\log_{10}[\Mbh/\Msun] = \{5,6,7,8\}$.      
    } 
    \label{fig_Ntde_time}
\end{figure}

\noindent
\emph{Time-dependent rates}: Using the above machinery, we can also compute the time-dependent rates of collisionless TDEs as the disk gradually grows in mass, say in accordance with model given in equation~\ref{disk_growth_model}. For this task, we numerically compute $\Ntde(\mu_i)$ using equation~\ref{Ntde} over grid points in disk mass $\{ \mu_i \}$ corresponding to a uniform time grid $\{t_i \}$ with intervals $\Delta t$. 
 The time interval $\Delta t$ should be long enough such that trapped orbits have time to librate through the island to reach the loss cone, which implies $ \Tlib \lesssim \Delta t \ll \Tgrow$. We choose $\Delta t = \TKep(\rh)$ to satisfies this condition, and then compute the \emph{instantaneous} rates of collisionless TDEs using discrete finite differences over this grid:  
 \beq  
\NdotCL(\mu_i) = \frac{ \Ntde(\mu_{i+1}) - \Ntde(\mu_{i-1}) }{2 \Delta t}
 \eeq 
Figure~\ref{fig_Ntde_time} shows the time-evolving rates of collisionless TDEs for an isotropic cluster ($\beta =0$) with density index $\gamma = 7/4$, considering a gas disk that grows to a maximum mass ratio of $\mu_0  =0.1$. For exponential disk growth, the collisionless TDE rate $\NdotCL$ continues increasing with time until the disk attains its maximum mass.  The collisionless TDE rate then vanishes as soon as disk enters a decay phase. Figure \ref{fig_Ntde_time} also compares these rates with collisional TDE rates, which are discussed later in this section. 

\begin{figure*}
    \centering
    \includegraphics[width=0.8\textwidth]{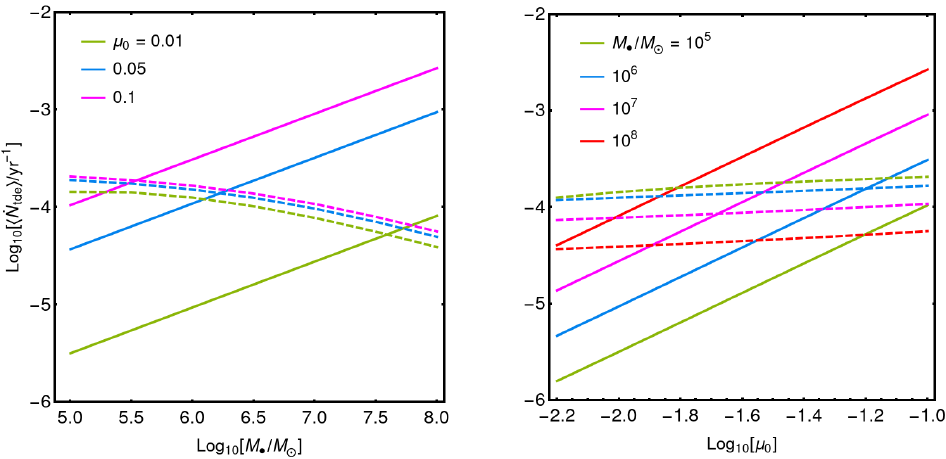}
    \caption{
    Comparisons of average collisionless TDE rates $\NdotCLav$ (solid lines) and collisional TDE rates $\NdotBB(\mu_0)$ (dashed lines). {\it Left panel:} TDE rates are plotted as functions of MBH mass $\Mbh$ for fixed disk-MBH mass ratio $\mu_0$ (in color). {\it Right panel:} TDE rates are shown as functions of maximum disk mass ratio $\mu_0$ for fixed $\Mbh$ (in color). Here we choose an isotropic cluster with $\beta =0$ and $\gamma = 7/4$. Collisionless TDE rates are notably dominant for high $\Mbh$ and $\mu_0$, in spite of a moderate enhancement in collisional rates owing to the loss wedge.         
    } 
    \label{fig_Ntde_compare}
\end{figure*}
 
 For analytical estimates, the instantaneous collisionless TDE rate can be defined as $\NdotCL(\mu) = \rmd \Ntde/\rmd \mu \times \dot{\mu} $. We can use the analytical form of $\Ntde(\mu)$ from equation~\ref{Ntde_approx1} and $\dot{\mu}$ from equation~\ref{disk_growth_model} to obtain the following explicit form of collisionless TDE rates as a function of time: 
 \begin{equation*} 
 \begin{split}
& \NdotCL(\mu; \mu_0 ) = \frac{(1- 2 \beta)}{2} \frac{\Ntde(\mu)}{\Tgrow(\mu_0)} \hspace{4cm} \\[1ex] %\frac{\mu_0}{\mu} \exp{\big( \frac{t}{\Tgrow(\mu_0)}\big)}  \\[1ex] 
& = \frac{(1- 2 \beta)}{2} \mathF(\beta,\gamma) \frac{N_{\rm h}}{\TKep(\rh)} \sqrt{\frac{\rtid}{\rh}} \, \mu_0 \, \mu^{\frac{1 - 2 \beta}{2}} 
\end{split}
\end{equation*} 
%\vspace{-0.3cm}
\beq 
\begin{split} 
&= \frac{(1- 2 \beta)}{2} \mathF(\beta,\gamma) \frac{N_{\rm h}}{\TKep(\rh)} \sqrt{\frac{\rtid}{\rh}} \, \mu_0^{\frac{3 - 2 \beta}{2}} \,   \exp{\bigg[ \frac{(1 - 2 \beta) t}{2 \, \Tgrow(\mu_0)} \bigg]} \\[1ex] 
& = 2 (1- 2 \beta) \NdotCLav \exp{\bigg[ \frac{(1 - 2 \beta) t}{2 \, \Tgrow(\mu_0)} \bigg]}.
  \end{split} 
  \label{NdotCL_time}
 \eeq 
 The dependence on $\beta$ inside the exponential highlights the fact that $\NdotCL$ varies over longer timescale $2 \Tgrow(\mu_0)/(1- 2 \beta)$ for higher $\beta$ (i.e. with a more radially biased velocity distribution).

\subsubsection{Comparison with Collisional TDE rates} 

\label{subsec_collisional_rates}

In the above calculation of collisionless TDE rates, we ignored the effects of angular momentum relaxation driven by 2B scatterings, which can also contribute to channelling stars into the TDE loss wedge during the disk lifetime. This is because the $\ell_z$-relaxation time $\Tlc \sim \Tbb \llc^2$ near the loss wedge boundary ($\ell_z = \llc$) can be shorter than the disk growth time $\Tgrow$ for some parameters of our interest. Collisional TDEs can contribute to net TDE flux during the disk lifetime, if $\Tlc < \Tgrow$. As before, we compare these timescales at $a = \rh$, and this condition translates to $\Mbh \, \rtid/(\mstar \, \rh) < \lnL/(2 \mu_0)$. This condition can be written more explicitly as $\Mbh \lesssim 6 \times 10^7 \Msun (\mu_0/0.1)^{-15/11}$ for Solar-type star and our fiducial form of $\rh$; MBHs smaller than this %critical mass will see collisional TDE rates dominate collisionless ones during AGN disk growth.   
value can see some contribution, as evaluated below, from collisional TDEs during AGN disk growth. We had checked towards the end of \S~\ref{sec_phy_setup} that collisional formation of TDEs would dominate over collisionless channel only for $\Mbh \lesssim 2 \times 10^6 \Msun$.   

%Earlier in \S~\ref{subsec_CL_TDEs}, we discussed the possible intervention of relaxation effects in the operation of the collisionless channel.
Here, we estimate the rates of collisional TDEs driven by relaxation due to 2B scatterings. We assume the extreme limit $\Tlc \ll \Tgrow$ for which the star cluster DF $F(a,j,\mu)$ relaxes near the loss wedge and adjusts fast enough to the slowly growing disk mass $\mu(t)$. Thus, the \emph{instantaneous} collisional TDE rate $\NdotBB$ can be approximated as steady state diffusion flux corresponding to a given $\mu$. We employ equations (72) and (73) of \citet{Vasiliev_Merritt2013} to evaluate $\NdotBB$ for an isotropic star cluster ($\beta = 0$):
\begin{equation*}
 \begin{split}
 & \NdotBB(\mu) = 4 \pi^3 (G \Mbh)^{3/2} N_{\rm h} \\[1ex]
 & \hspace{2.4cm}\int_0^{\rh} \rmd a \; \frac{\sqrt{a}  F_0(a)}{ \Tbb(a) \left[ \alpaxi + \log{( 1 / \Reff )}  \right] }   
\end{split}
\end{equation*}
with parameter:
\beq 
\begin{split} 
& \alpaxi = \begin{cases}
    \qs/\qaxi \quad , \quad \qaxi < 1 \\ 
    \qs \quad , \quad \qaxi \geq 1 \\
\end{cases} 
\end{split}
\label{Ndot_BB_mu}
\eeq 
\begin{equation*} 
 \mbox{and }\; \qaxi = \TKep \, \js/(\Tlib \, \llc) \quad, \; \qs = \TKep/(\Tbb \, \llc^2)
\end{equation*}  
where $F_0(a) = F_0(a,1)$ from equation~\ref{F0_simple}. Since the action $\js$ of the SO and its associated libration timescale $\Tlib$ have a negligible dependence on $\ell_z \in [-\llc,\llc]$ (see equation~\ref{jco_approx}), we choose $\ell_z= \llc$ to evaluate these quantities numerically; see appendix~\ref{app_actions} for details. The effective size of the loss region is $\Reff = \js^2(0.1 + 0.9 \, \llc/\js)$ (compared to loss cone size $\sim \llc^2$ for spherical systems). Due to its logarithmic dependence, this extended loss cone (the so-called loss wedge), can enhance the rates of collisional TDEs only by a factor of a few ($\sim 2-3$) in axisymmetric systems, compared to spherical ones \citep{Vasiliev_Merritt2013}. However, for the full loss cone limit with $\alpaxi = \qs \gg 1$, 2B TDE rates become nearly equal for both spherical and axisymmetric geometries.

We calculate $\NdotBB(\mu)$ for various values of $\Mbh = 10^{5-8} \Msun$ and $\gamma = 7/4$ by numerically evaluating the integral of equation~\ref{Ndot_BB_mu}, and show these results as function of time in the figure~\ref{fig_Ntde_time}, for our chosen disk growth model $\mu(t)$ of equation~\ref{disk_growth_model}. Unlike the collisionless case, collisional TDE rates are decreasing function of MBH mass, with $\NdotBB \propto N_{\rm h}/\Tbb(\rh) \propto \Mbh^{-2/5}$. Further, $\NdotBB \propto 1/\log{(1/\mu)}$ is a more slowly increasing function of $\mu$ (in comparison to $\NdotCL(\mu) \propto \sqrt{\mu}$ from equation~\ref{NdotCL_time}), owing to logarithmic dependence on $\Reff \propto \js^2 \propto \mu$. The figure~\ref{fig_Ntde_time} (for $\mu_0=0.1$) highlights the importance of collisionless contribution to TDE rates for high $\Mbh$. The ratio $\NdotCL/\NdotBB \gtrsim 10$ for $\Mbh \gtrsim 10^7 \Msun$ for at least a few disk growth timescales $\Tgrow$. On the other hand, for low $\Mbh \lesssim 10^5 \Msun$, collisionless rates are always sub-dominant with $\NdotCL \lesssim \NdotBB$. 

The above results are further quantified in figure~\ref{fig_Ntde_compare}, which compares average collisionless TDE rates $\NdotCLav$ and 2B TDE rates $\NdotBB(\mu_0)$ as functions of $\Mbh$ and $\mu_0$. For a typical $\mu_0 = 0.1$, the relative rate enhancement spans a wide range for $\Mbh = 10^{6-8} \Msun$, with $\NdotCLav/\NdotBB \simeq 2 - 47$. %For a simplified analysis, we consider here an average value of $\NdotCLav/\NdotBB \simeq 8$, as is attained for $\Mbh = 10^7 \Msun$.
The TDE rate in AGN is poorly understood at present, so it is not possible to definitively test this prediction.  However, recent observational samples of nuclear infrared (IR) flares may offer a handle on this elusive transient population.  

Infrared dust echoes have been observed to follow standard TDEs in vacuum galactic nuclei \citep{vanVelzen+16, Jiang+16, Dou+16}.  While not all standard TDEs produce IR dust echoes, roughly $\approx 30\%$ do so \citep{Jiang_2021}.  Recently, \citet{vanVelzen24} identified a population of high-amplitude IR flares in AGN whose properties matched those of IR dust echoes to standard TDEs.  Although some members of this population could represent extreme AGN variability, the complete absence of these flares from SMBHs above the Hills mass strongly suggests a TDE origin \citep{vanVelzen24}.

If we assume that a fraction $\fagn \approx 0.05$ of all galaxies are type 1 AGN \citep{Pimbblet_2013,Oh_2015,Lopes_2017} suitable for detection of large nuclear flares (e.g. collisionless TDEs), the expected ratio of observed TDE detections $\Nclo/\Nbbo \simeq \fagn \NdotCLav/\NdotBB \simeq 0.1-2.4$, where $\Nclo$ and $\Nbbo$ are the number of collisionless and collisional TDEs observed during a given time-period. %Recent observations \citep{vanVelzen24} suggest that $\sim$6 TDE detections occur for every 18 large AGN flares, detected from dust echoes in IR. Given that $\sim 0.3$ of the optical TDEs were recovered by this method \citep{Jiang_2021,vanVelzen24}, 
The observationally inferred ratio $\Nclo/\Nbbo \approx 0.9$, under the simplifying assumptions that (i) all large AGN flares in \citet{vanVelzen24} correspond to AGN TDEs, (ii) all AGN TDEs have detectable IR echoes, and (iii) only $30\%$ of standard TDEs have such IR echoes\footnote{These tentative observational indications of an enhancement of TDE rates in AGN were communicated to us by Dr.~Sjoert van Velzen.}. Turning this calculation around, it suggests a rate enhancement of TDEs in AGN (compared to vacuum galactic nuclei) of $\NdotCL/\NdotBB \sim 20$.  This \emph{very} rough observational estimate is consistent with our theoretical predictions, but more thorough comparisons are needed in the future to better investigate the question of TDE rates in AGN.  In particular, such efforts would need a clear framework to distinguish AGN TDEs from extreme AGN variability \citep{Zabludoff_2021}, as well as to account for any change in the luminosity function between classical TDEs and those occuring in AGN.

\section{Discussion and Conclusions}

\label{sec_discus}

In this work, we investigated a collisionless channel for the formation of TDEs during an AGN episode. Our theoretical calculations are motivated by recently discovered TDE candidates in AGN and gas rich nuclei \citep{Masterson_2024}, which have often been ignored in earlier optical/UV transient surveys. As the overall number of TDEs is expected to dramatically increase in the coming years \citep{Bricman_2020, Shvartzvald_2023}, it will become important to carefully investigate occurrence rates and observational signatures of TDEs arising in diverse nuclear environments. The novel collisionless channel that we proposed invokes a time-evolving AGN disk, and results in much higher TDE rates than those examined in earlier studies considering a static axisymmetric perturbation \citep{Magorrian_Tremaine_1999,Vasiliev_Merritt2013}. Due to the growth of the disk over a time $\Tgrow \sim$ few $\times$ Myrs, the loss wedge expands to capture high-$\ell$ but low-$\ell_{\rm z}$ stars into the libration island; these can become TDEs during their libration cycles, over a time $\Tlib \ll \Tgrow$. Since the stars are fed to the loss wedge over a time $\Tgrow$, this channel leads to much higher collisionless TDE rates ($\NdotCL$) compared to the collisional rates $\NdotBB$ predicted either into a spherically symmetric loss cone \citep{Wang_Merritt_2004, Stone_Metzer_2016}, or into an axisymmetric loss wedge \citep{Magorrian_Tremaine_1999}, as  $\Tbb \gg \Tgrow$ generally. 

We demonstrated this effect for an initially spherical star cluster perturbed by an adiabatically evolving gas disk.  Working within the region of influence  ($a < \rh$) of the central MBH, the regime of secular dynamics was well justified. %, as was the approximation of a nearly Keplerian potential. 
We utilized effective adiabatic conservation of actions to evaluate the time-dependent, non-linear evolution of the stellar DF $F(a,j,\mu)$. This allowed us to track the rate of stellar capture into the loss wedge, and to numerically evaluate the resulting collisionless TDE rates. The underlying simplicity of loss wedge dynamics also allowed us to compute approximate analytical expressions for rates of collisionless TDEs (equations \ref{Ntde_approx2}, \ref{NdotCLav_approx}, and \ref{NdotCL_time}).

The resulting collisionless TDE rates in galactic nuclei with a growing AGN can be much higher than those from the traditional channel (driven by 2B scatterings), especially for more massive MBHs $\Mbh$ and larger fractional disk mass $\mu_0$. For $\Mbh = 10^7 \Msun$ and $\mu_0 = 0.1$, the average collisionless rate $\NdotCLav$ is higher than $\NdotBB$ by about an order of magnitude (for an isotropic velocity distribution; $\beta = 0$). Our results indicate that $\NdotCLav \propto \Mbh^{7/15} \mu_0^{(3 - 2 \beta)/2}$, where $\beta$ is the standard anisotropy parameter characterising the distribution of stellar velocities. Our numerical evaluations of $\NdotCLav$ produced an exponential rise with $\beta$ (i.e. increasingly radially biased stellar velocities\footnote{This result recalls the similar sensitivity of collisional TDEs to $\beta$ \citep{LezhninVasiliev15, Stone_2018}.}), with $\NdotCLav$ spanning two orders of magnitude for $\beta \in [-1,1/2]$. Unlike collisional TDE rates, we find a very weak dependence of $\NdotCLav$ on the density power-law index $\gamma$ characterizing the slope of the stellar distribution. Interestingly, observational evidence tentatively suggests a rate enhancement $\NdotCL/\NdotBB \sim 20$ \citep{vanVelzen24}, in line with our theoretical calculation (though we caution that the inferred rate of AGN TDEs is highly uncertain).

Our current exploration of the subject is far from complete and future investigations are necessary to explore the complete dynamics of NSCs hosting time-evolving AGN disks. We have evaluated collisionless TDE rates in an empty loss wedge regime without accounting for the effects of 2B scatterings self-consistently, which might become important for MBHs with $\Mbh \lesssim 10^6 \Msun$. Furthermore, disk formation around an MBH in real astrophysical settings can be quite messy, and non-axisymmetric geometries are plausible, especially in the limit of chaotic accretion \citep{KingPringle06}.  A globally eccentric accretion disk may enhance TDE rates due to additional secular effects not explored here \citep{Madigan_2018}. 

Perhaps most importantly, our calculation does not self-consistently include dissipation effects arising from gas dynamical friction and geometric drag \citep{Artymowicz1993}. Our simple analysis (appendix~\ref{app_gas_drag}) of the involved timescales suggests that gas dynamics will only have a minimal influence on collisionless TDE rates in a growing AGN disk, as most of these TDEs are sourced from semimajor axes near $r_{\rm h}$ where the gas drag timescales are quite long (also in agreement with previous analysis by \citet{MacLeod_2020} for collisional TDEs). However, while this paper was being completed, \citet{Wang+24b} published a study investigating the way in which gas drag can produce AGN TDEs.  Aside from a relatively small population of ``disk TDEs'' (TDEs from stars whose initial orbits are by chance retrograde and co-aligned with the plane of the AGN), \citet{Wang+24b} find that most of the TDEs produced by gas drag are first aligned into the disk, and then circularize their orbits and migrate inwards.  While significant AGN variability may result from the eventual tidal interaction of these stars with the central MBH, these low-eccentricity and low-inclination events are quite distinct from the high-eccentricity, high-inclination TDEs studied in our work.

While we defer these directions of exploration to a future study, we expect our qualitative conclusions to be reasonably robust.  Time-dependent quadrupole moments produced by growing AGN disks can dramatically increase TDE rates in galactic nuclei.  This effect is largest for the highest-mass MBHs capable of producing TDEs (i.e. those near the Hills mass). 
Understanding the observational signatures of such AGN TDEs is an important challenge for the rapidly growing field of nuclear transients.

\begin{acknowledgments}

Both KK and NCS gratefully acknowledge support from the Israel Science Foundation (Individual Research Grant 2565/19). NCS acknowledges additional support from the Binational Science Foundation (grant Nos. 2019772 and 2020397). Both authors thank Edwin Chan, Tsvi Piran, Aleksey Generozov,  Eugene Vasiliev and Brian Metzger for helpful discussions on the manuscript, and are particularly grateful to Sjoert van Velzen for his suggestions concerning tentative observational evidence for TDE rate enhancements in AGN.

\end{acknowledgments}

\bibliography{TDE_refs}{}
\bibliographystyle{aasjournal}

\appendix

%\FloatBarrier

\section{Axisymmetric Potential-Density Pairs} 
\label{app_axis_pot_den}

The power-law forms of axisymmetric density $\rhoaxis(r,\theta)$ and potential $\Phiaxis(r,\theta)$,
\beq
\rhoaxis(r ,\theta) = \rho_0 \bigg(\frac{\rh}{r} \bigg)^{\!\!\gaxis} S(\theta) \qquad , \qquad \Phiaxis(r, \theta) = 4 \pi G \, \rho_0 \, \rh^{\gaxis} \, r^{2-\gaxis} \, \Psi(\theta) 
\eeq 
satisfy the following angular Poisson equation,
\beq  
\frac{1}{s_{\theta}} \frac{\rmd}{\rmd \theta} \bigg( s_{\theta} \frac{\rmd \Psi}{\rmd \theta}  \bigg) +
(2-\gaxis) (3-\gaxis) \Psi = S,
\label{poisson_ang}
\eeq 
where $s_{x} = \sin{x}$ and $c_{x} = \cos{x}$ throughout this section.

\noindent
We choose an angular potential of the form,
\beq 
\Psi(\theta) = \psi_0 + \psi_1 \, |c_{\theta}| + \psi_2 \, c_{\theta}^2
\eeq 
where $ \psi_{0,1,2} $ are constants, to be determined later, that can yield disk densities of interest. The $|c_{\theta}|$ term gives an infinitesimally thin component of disk density. Substituting the above form of $\Psi$ into equation~(\ref{poisson_ang}) gives the angular density function:
\beq 
S(\theta) = 2 \psi_1 \, \delta{\bigg( \theta - \frac{\pi}{2} \bigg)} + \left[ t_{\gamma} \psi_0 + 2 \psi_2 + \left( t_{\gamma} -2    \right) \psi_1 \, |c_{\theta}| + \left(     t_{\gamma}  -6 \right) \psi_2 \, c_{\theta}^2 \right]  
 \eeq 
where $t_{\gamma} = (2-\gaxis)(3-\gaxis)$. The first term corresponds to an infinitesimally thin component of density, and the term inside the ``$\left[ \quad \right]$'' brackets corresponds to an envelope above and below the disk plane. By demanding a particular functional form of the envelope density $\propto (1 - |c_{\theta}|)^2$, which decreases monotonically with increasing latitudes from disk plane, we obtain:% $\psi_0/\psi_1$ and $\psi_2/\psi_1$:
\beq 
 \psi_0/\psi_1 = - \frac{(t_{\gamma}-2) (8-t_{\gamma})}{2 t_{\gamma} (6 - t_{\gamma})}  \quad ; \quad 
 \psi_2/\psi_1 = \frac{t_{\gamma}-2}{2 (6 - t_{\gamma})} \\[1ex] ,
\eeq 
yielding an angular potential $\Psi(\theta)$ and the following form of angular density $S(\theta)$:
\beq 
S(\theta) =  2 \psi_1 \, \delta \bigg( \theta - \frac{\pi}{2} \bigg) + \frac{(2- t_{\gamma})\psi_1}{2} \left( 1 - |c_{\theta}|  \right)^2 .
\eeq 

\noindent
For a total disk mass $\Md$ inside radius $\rh$ (integrating over $\theta \in \left[0, \pi \right]$), we have: 
\beq 
\rho_0 \, \psi_1 = \frac{3 \Md}{2 \pi  \rh^3} \frac{ 3 - \gaxis}{ 8 -t_{\gamma}}
\eeq 
and the final form of an axisymmetric potential-density pair for a disk with a general radial density index $\gaxis$:
\beq 
\begin{split} 
& \rhoaxis(r ,\theta) = \frac{3 \Md}{2 \pi  \rh^3} \frac{ 3 - \gaxis}{ 8 -t_{\gamma}} \bigg(\frac{\rh}{r} \bigg)^{\!\!\gaxis}  \bigg\{ 2 \, \delta \bigg( \theta - \frac{\pi}{2} \bigg) + \frac{(2- t_{\gamma})}{2} \left( 1 - |c_{\theta}|  \right)^2   \bigg\}  \\[1ex] 
&  \Phiaxis(r, \theta) =  \frac{6 G \Md}{  \rh} \frac{ 3 - \gaxis}{ 8 -t_{\gamma}} \,  {\bigg( \frac{r}{\rh}  \bigg)}^{2-\gaxis} \, \bigg\{  - \frac{(t_{\gamma}-2) (8-t_{\gamma})}{2 t_{\gamma} (6 - t_{\gamma})}  +  |c_{\theta}| + \frac{t_{\gamma}-2}{2 (6 - t_{\gamma})} \, c_{\theta}^2    \bigg\}
\end{split} 
\eeq 
We use $\gaxis = 3/2$, which gives an analytical form of the orbit-averaged potential (easing further calculations in the paper), and which is also in reasonable alignment with both very simple models for AGN disks as well as those which self-regulate due to accretion feedback from stellar-mass compact objects \citep{GilbaumStone22}.  We note, however, that AGN disks which self-regulate due to feedback from stellar processes can achieve a much steeper density profile \citep{SirkoGoodman03, Thompson+05}. 

\section{Transformation to action-angles and Orbit-averaging} 
\label{app_orbit_av_Calci}

The physical space variables $\{r,\theta \}$ can be transformed to the actions or Keplerian orbital elements through the following expressions \citep{Murray_Dermott1999book,Sambhus2000}: 
\begin{subequations}
\beq 
r = a  (1- e \cos{\eta})
\eeq 
\beq 
\cos{\theta} = \frac{z}{r} = \frac{\sin{i} ( \sin{g} (\cos{\eta}-e) + \cos{g} \sqrt{1-e^2} \sin{\eta} )  }{1 - e \cos{\eta}}
\eeq  
\label{tras_r_theta}
\end{subequations}
where $\eta$ is the eccentric anomaly. 

For the orbit-averaging disk potential $\phid$, we need to evaluate three types of integrals $\left< \sqrt{r} \right>$, $\left<\sqrt{r} \cos^2{\theta}\right>$ and $\left< \sqrt{r} |\cos{\theta}|  \right>$; where $\left<`` \;" \right> \equiv (2\pi)^{-1} \oint \rmd w \, `` \; "  = (2\pi)^{-1} \oint \rmd \eta \; (1 - e \cos{\eta}) \, `` \; "  $. We use $w = \eta - e \sin \eta$. The three integrals can be expressed in terms of elliptic integrals with the standard definitions given below. % in equations~(A2) and (A3) of \citet{Kaur_2018cusp}.  
\beq 
\mathF(\xi_0,k) = \int_{0}^{\xi_0} \rmd \xi \; \frac{1}{\sqrt{1 - k^2 \sin^2{\xi}}}  \qquad \; \qquad \mathK(k) = \int_{0}^{\pi/2} \rmd \xi \; \frac{1}{\sqrt{1 - k^2 \sin^2{\xi}}} 
\eeq 
are the incomplete and complete elliptic integrals of first kind. 
\beq 
\mathE(\xi_0,k) = \int_{0}^{\xi_0} \rmd \xi \; {\sqrt{1 - k^2 \sin^2{\xi}}}  \qquad \; \qquad \mathE(k) = \int_{0}^{\pi/2} \rmd \xi \; {\sqrt{1 - k^2 \sin^2{\xi}}} 
\eeq 
are the incomplete and complete elliptic integrals of second kind. 

The first integral is straightforward to solve giving:
\beq 
\left< \sqrt{r} \right> = \frac{2 \sqrt{a (1+e)}}{3 \pi} \big( 4 \, \mathE(k) - (1-e) \, \mathK(k) \big)  
\eeq 
where $k = \sqrt{2 e/ (1+e)}$. Similarly, the second integral can be solved to give: 
\beq 
\left<\sqrt{r} \cos^2{\theta}\right> = \frac{ \sin^2{i} \sqrt{a (1+e)}}{3 \pi} \bigg[ 4 \mathE(k) - (1 - e) \mathK{(k)}   + \frac{ \cos{(2 g)} }{e^2} \big\{ 4 (1 - 2 e^2) \mathE(k) -(1-e)(4-5 e^2) \mathK(k) \big\}   \bigg] 
\eeq 

The below description details the evaluation of the third integral: 
\beq 
\left< \sqrt{r} |\cos{\theta}|  \right> = \frac{\sqrt{a} \sin{i} }{2 \pi} \int_0^{2\pi} \rmd  \eta \; (1 - e \cos{\eta})^{1/2}  
  \left| \sin{g} (\cos{\eta}-e) + \cos{g} \sqrt{1-e^2} \sin{\eta}   \right|
\eeq 
We can express $\left| \sin{g} (\cos{\eta}-e) + \cos{g} \sqrt{1-e^2} \sin{\eta}   \right| = \sqrt{1 - e^2 \cos^2{g}}  
\left|  \cos{(\eta - \eta_0)} - \cos{\theta_0}   \right|$, where:
\beq 
\eta_0(e,g) = \tan^{-1}(\sqrt{1-e^2} \cot{g}) \qquad , \qquad \theta_0(e,g) = \tan^{-1}{\bigg(  \frac{\sqrt{1-e^2}}{ e \sin{g} }  \bigg)  } 
\eeq 
Following the approach of \citet{Kaur_2018cusp} (their appendix A), we can rewrite the third integral as:  
\beq 
\begin{split}
 \left< \sqrt{r} |\cos{\theta}|  \right> & = \frac{\sqrt{a} \sin{i} }{2 \pi} \bigg|  \int_0^{2\pi} \rmd  \eta \; (1 - e \cos{\eta})^{1/2}  
  \left( \sin{g} (\cos{\eta}-e) + \cos{g} \sqrt{1-e^2} \sin{\eta}   \right)  \\[1ex]
& \qquad  - 2  \int_{\eta_0 + \theta_0}^{2\pi + \eta_0 - \theta_0} \rmd  \eta \; (1 - e \cos{\eta})^{1/2}  
  \left( \sin{g} (\cos{\eta}-e) + \cos{g} \sqrt{1-e^2} \sin{\eta}   \right) 
  \bigg|
\end{split}
\eeq 
We find: 
\beq 
\left< \sqrt{r} |\cos{\theta}|  \right> = \frac{2 \sqrt{a(1+e)} \sin{i} }{ 3 \pi e  } 
\bigg[ \sqrt{1-e}  \cos{g} \{ ( 1 + e \cos{(2 \eta_1)} )^{3/2} - ( 1 + e \cos{(2 \eta_2)} )^{3/2}    \} + A(e,g) \sin{g}  \bigg]
\eeq 
where 
\beq 
\begin{split}
A(e,g)  = (1-e) \big\{ & \mathK(k) +\mathF(\eta_1,k) - \mathF(\eta_2,k)  \big\} - (1+3 e^2) \big\{ \mathE(k) + \mathE(\eta_1,k) - \mathE(\eta_2,k)   \big\} \\[1ex]
\qquad \quad & + e \sin{(2 \eta_2)} \sqrt{\frac{1+e \cos{(2 \eta_2)}}{1+e}} 
- e \sin{(2 \eta_1)} \sqrt{\frac{1+e \cos{(2 \eta_1)}}{1+e}}
\end{split} 
\eeq 
and 
\beq 
\eta_1(e,g) = \frac{\eta_0 + \theta_0 - \pi}{2} \quad ; \quad \eta_2(e,g) = \frac{\eta_0 - \theta_0 + \pi}{2}
\eeq 

Using above expressions, we have the following exact form of averaged disk potential:
\beq 
\begin{split}
\Phid &=  \frac{ 12 G \Md  }{29 \pi \rh} \bigg(\frac{a}{\rh} \bigg)^{1/2} \sqrt{(1+e)} \bigg[ 
\frac{145}{63}   \big( 4 \, \mathE(k) - (1-e) \, \mathK(k) \big)   \\[1ex]
& - \frac{5}{42}  \sin^2{i}  \bigg\{   4 \mathE(k) - (1 - e) \mathK{(k)}   + \frac{ \cos{(2 g)} }{e^2} \big\{ 4 (1 - 2 e^2) \mathE(k) -(1-e)(4-5 e^2) \mathK(k) \big\}    \bigg\} \\[1ex]
& + 
\frac{2  \sin{i} }{  e  } 
\bigg\{ \sqrt{1-e}  \cos{g} \{ ( 1 + e \cos{(2 \eta_1)} )^{3/2} - ( 1 + e \cos{(2 \eta_2)} )^{3/2}    \} + A(e,g) \sin{g}  \bigg\}
\bigg] 
\end{split}
\label{Phid_exact}
\eeq 
We approximate the above potential to the following form, which is accurate upto $\simeq 4$ per cent.  
\beq  
 \Phid \simeq  \frac{  G \Md  }{ \rh} \bigg(\frac{a}{\rh} \bigg)^{1/2} \bigg\{
 T_1(e)  + T_2(e,g) \sin^2{i}  + T_3(e,g) \sin{i}    \big\}  
 \bigg\} 
 \label{Phid_approx} 
\eeq 
where functions $T_1(e)$, $T_2(e,g)$ and $T_3(e,g)$ are given below: 
\beq 
\begin{split} 
& T_1(e) = \frac{10}{7} (1 + 0.2 e^2) \\[1ex]
& T_2(e,g) = - \frac{30}{203}  \{ 0.5 + 0.1 e^2 - 0.6 e^2 \cos{(2 g)} \} \\[1ex] 
& T_3(e,g) = \frac{36}{29} \big\{ 0.637 \sqrt{1-e^2} + 1.18 e^2 (1 - \sqrt{1-e^2}) |\sin{g}| + 0.940 e^2 \sqrt{1-e^2} \sin^2{g}  \big\} 
\end{split}
\label{T_val}
\eeq 

\section{Secular dynamics - evaluation of actions} 
\label{app_actions}

For our semi-analytical calculations, we employ a normalized secular Hamiltonian $H=   (\Phic + \Phid)/\mathcal{H}$, where the normalization factor $\mathcal{H}$ is defined as follows: 
\beq 
\mathcal{H} = \frac{G \Mbh}{r_h} \frac{\algam}{2 - \gamma} \bigg( \frac{a}{r_h}  \bigg)^{2-\gamma}  \,.
\label{H_norm}
\eeq 
The resulting Hamiltonian is explicitly given as (for the choice of approximate disk potential $\Phid$ of equation~\ref{Phid_approx}):
\beq 
H(\ell,g; \lz,\chi) = - \ell^2 + \chi \bigg\{  T_1(e)  + T_2(e,g) \sin^2{i}  + T_3(e,g) \sin{i}  \bigg\}
\label{H_explicit}
\eeq 
where we drop some terms dependent on only $a$, that remains a constant of motion in secular dynamics. The perturbation strength $\chi$ (given in equation~\ref{chi_def}) is also given below for convenience:
\begin{equation*} 
\chi = \mu \frac{2 - \gamma}{\algam} \bigg( \frac{a}{\rh} \bigg)^{\gamma - 3/2} \,. 
 \end{equation*} 
Because of the above normalization of Hamiltonian and also, the use of normalized angular momentum $\ell = L/I$, we need to employ a normalized time $\tau = (\mathcal{H}/I) \, t $ given explicitly as:
\beq 
\tau = \frac{\algam}{2-\gamma} \bigg(\frac{a}{\rh} \bigg)^{3/2-\gamma} \frac{2 \pi  \, t}{\TKep(\rh)}
\label{tau_def}
\eeq 
where $t$ is the physical time. The Hamiltonian equations of motion are then given as:
\beq 
\frac{\rmd g }{\rmd \tau} = \frac{\p H}{\p \ell} \qquad \qquad , \qquad \qquad \frac{\rmd \ell }{\rmd \tau} = - \frac{\p H}{\p g}
\label{eom}
\eeq 
which define the coupled evolution of $\ell$ and $g$ (figure~\ref{fig_Hcont_3reg}). The dynamics also includes nodal precession of the orbital plane defined by ${\rmd h}/{\rmd \tau} = {\p H}/{\p \ell_z}$, but it remains unimportant for our purposes. As seen earlier in section~\ref{sec_phy_setup}, $\ell_z$ is the integral of motion. With the quasi-static assumption valid over timescale $t \ll \Tgrow$, $\chi$ can be assumed as a constant, and $H$ is then the second integral of motion. 

\medskip 

\noindent
\emph{Fixed points}: We have the two sets of hyperbolic and elliptic fixed points in $\{g,\ell\}$-plane, as shown in the figure~\ref{fig_Hcont_3reg}. The hyperbolic fixed points $\ell = \lhyp$ at $g = 0^{\circ}, \, 180^{\circ}$, can be located by using $\p H/\p \ell =0$ (a linear polynomial equation in $\lhyp^2$), which gives the following analytical form:
\beq 
\lhyp = \sqrt{\lz^2 +  \displaystyle {\frac{0.156}{\big( 0.360 + \frac{1}{\chi}    \big)^2} }  }  \,.
\eeq 
Similarly, elliptic fixed points $\lell$ at $g = 90^{\circ}, \, 180^{\circ}$ are located by $\p H/\p \ell =0$, which is a high-order polynomial equation in $\lell$ and is  %l^12
solved numerically for a root $\lell \in (\lz, 1 ) $.

\medskip 

\noindent
\emph{Spatial extent of orbits}: We can identify the separatrix orbit (SO) with contour value $\Hs = H(\lhyp,0^{\circ}; \lz,\chi)$, and use it to distinguish between circulating orbits (COs) with $H < \Hs$  and librating orbits (LOs) with $H > \Hs$. For upper COs, we locate the points of  minimum and maximum $\ell$ denoted as $\lmin$ and $\lmax$ by numerically solving for the intersections of $H$-contour with $g =0^{\circ}$ and $90^{\circ}$. For lower COs, the maximum and minimum points are reversed at $g = 0^{\circ}$ and $90^{\circ}$. Similarly, for LOs, both $\lmin$ and $\lmax$ are the two points of intersection of $H$-contour with $g = 90^{\circ}$. 

We determine the angular extent of a LO in $g$, defined by minimum and maximum periapsidal angles $\gmin$ and $\gmax$ through the procedure described here. A major advantage of using the form of $H$ given in equation~\ref{H_explicit} is that it has a quadratic form in $|\sin{g}|$. Hence, we can describe the $H$-contours with an analytical form of $g(\ell; H,\lz,\chi)$, and locate the $\gmin < 90^{\circ}$ as the extremum point with $\rmd g(\ell; H,\lz,\chi)/\rmd \ell = 0$ and the maximum point $\gmax = 180^{\circ}-\gmin$. 

\medskip 

\noindent
\emph{Actions}: Since the action $\jc$ for a CO equals the area under the corresponding $H$ contour in $\{g,\ell\}$-plane, with $\jc = (180^{\circ})^{-1} \int_{0^{\circ}}^{180^{\circ}} \ell(H; \mu, a, \ell_z) \, \rmd g $. It is sufficient for our calculation to use an approximate measure of this action $\jc \simeq (\lmin + \lmax)/2$. Also, the action $\jl$ for a LO equals the area bounded inside the corresponding closed $H$-contour with $\jl = (180^{\circ})^{-1} \oint \ell(H; \mu, a, \ell_z) \, \rmd g $, which we approximate as:
\begin{equation*}
\jl \simeq  \frac{\lmax - \lmin}{2} \, \frac{\gmax - \gmin}{180^{\circ}} \, .
\end{equation*}
These approximate forms of actions are also valid at the boundary of libration island, so that the action $\js$ for SO satisfies the relation $\js = \jco - \jco'$, where $\jco$ and $\jco'$ are the actions for the upper and lower COs just touching the SO.

\section{Dissipation effects due to gas disk} 
\label{app_gas_drag}

%Some earlier works that address this problem \citep{Rein_2012,Generozov_2023,Nasim_2023}.

In the earlier sections, we neglected the effect of gas dissipation on stellar orbits \citep{Artymowicz1993}, while accounting for only gravitational potential of the gas disk. As a star, following an orbit with finite inclination $i > 0^{\circ}$, passes through the disk near the nodal points $r_n$, it experiences a dissipative force directed opposite to the relative velocity $\vrelvec$ of the star with respect to the local Keplerian velocity $\vg = \sqrt{G \Mbh/r_n}$ of gas. Here we assume that the stellar orbit lies outside the vertical extent of the gas disk during most of the time, and experience dissipative force only for a fraction $\Sigd/(\rhod |v_z| \TKep)$ of its Keplerian orbit as it crosses through the disk near the nodes \footnote{We evaluate the force only at the nodal point and extrapolate its action throughout the entire vertical extent of the disk measured by $\Sigd/\rhod$.}. The orbit-averaged dissipative force (contributed near a node $r_n$\footnote{Similar contribution comes from another node too, that is in a different direction depending on the local relative velocity.}) that acts on a unit mass :
\beq  
\Fdisvec = - \frac{\Sigd(r_n)}{\mstar |\vstarz| \TKep(a)} \bigg\{ \frac{4 \pi \Cdf G^2 \mstar^2 }{\vrel^2} + \pi \Rstar^2 \vrel^2  \bigg\} \, \frac{\vrelvec}{\vrel}
\eeq 
where $\vrel$ is the magnitude of relative velocity and $\vstarz$ is the vertical velocity of the star at the node. The first term in the bracket refers to the local gas dynamical friction \citep{Ostriker_1999} and the second term represents the hydrodynamical drag. The dominance of either of the two phenomena depends on the comparative strength of relative velocity $\vrel$ and escape speed $\vesc = \sqrt{2 G \mstar/\Rstar}$ from the stellar surface. For higher relative velocity $\vrel$, the drag term due to the headwind hitting at the surface of star dominates. On the contrary for lower $\vrel$, dynamical friction term dominates as the star gets sufficient time to accumulate gas in a trailing wake that exerts a strong frictional force. Following earlier works \citep{Generozov_2023}, we take the dynamical friction constant $\Cdf = 3$. Also, the gas disk surface density $\Sigd(R) = 0.764 (\Md/\pi \rh^2 ) \sqrt{\rh/R}$ for disk potential-density pair given by equation~\ref{disk_den_pot}. For highly eccentric orbits with line of apses roughly aligned with the disk, the relative velocity $\vrel \sim \vstar$ at the inner node (close to periapsis) is very high, leading to dominant contribution from the geometric drag. In contrast, at the outer node (close to apoapsis) generally gas dynamical friction tends to dominate. 

The net effect of this dissipative force on a stellar orbit depends on the direction of the force relative to orbital plane, and hence indirectly on the orientation of stellar orbit with respect to disk plane. We define the coordinates $\{r, \phi_p , z_p \}$ such that $\zcap_p$ is directed along the angular momentum of stellar orbit, and $\phicap_p$ is the azimuthal angle in its orbital plane, measured with respect to line of nodes such that $\phi_p = 0^{\circ} $ at the ascending node $N_1$ (and $\phi_p = 180^{\circ}$ at the descending node $N_2$) \footnote{$\phi_p$ can be related to the free anomaly $f$ of the orbit as $f = \phi_p - g$.}. Also, $r$ is the usual radial distance from MBH. Let $\{F_r, F_{\phi p} , F_{z p} \}$ are the radial, tangential and normal components of the dissipative force $\Fdisvec$ in these coordinates. For evolution in $a$, only $F_r$ and $F_{\phi p}$ contribute \citep{Burns_1976,Murray_Dermott1999book}, so that: 
\beq 
\frac{\rmd a}{\rmd t} = \frac{2 a^{3/2}}{\sqrt{G \Mbh (1-e^2)}} \big[  F_r \, e \sin{(\phi_p - g)} + F_{\phi p} \, (1 + e \cos{(\phi_p - g)})     \big] \,.
\eeq 
We estimate the migration timescale $T_a$ over which semi-major axis $a$ of a Keplerian stellar orbit shrinks, as follows:
\beq 
T_a = \frac{a}{|\rmd a/\rmd t|} 
\eeq 
We evaluate this timescale for a given orbit $\{a,e,i,g\}$\footnote{Owing to axisymmetry of the problem, $T_a$ is independent of argument of ascending node $h$.} at both nodes with $\phi_p = 0^{\circ} ,\, 180^{\circ}$ and choose the minimum value. We evaluate this timescale along the separatrix which represents the secular orbit for most collisionless TDEs, and compare it with the timescale $\Tlib$ to librate to the loss cone $\ell = \llc$. When this phase dependent libration time becomes less than $\TKep(a)$ (near $\llc$), we use $\TKep$ for comparison with the migration time $T_a$, as $\TKep$ is a rough representative of the dynamical time over which a star inside/on the loss cone would get disrupted. In other words, we use the ratio $T_a/{\rm max}[\TKep,\Tlib]$ to judge the significance of dissipation for a prospective TDE. If this ratio is much greater than unity, then these dissipative effects are not important and should not hinder the formation of collisionless TDE under consideration.  

\begin{figure*}
    \centering
    \includegraphics[width=0.95\textwidth]{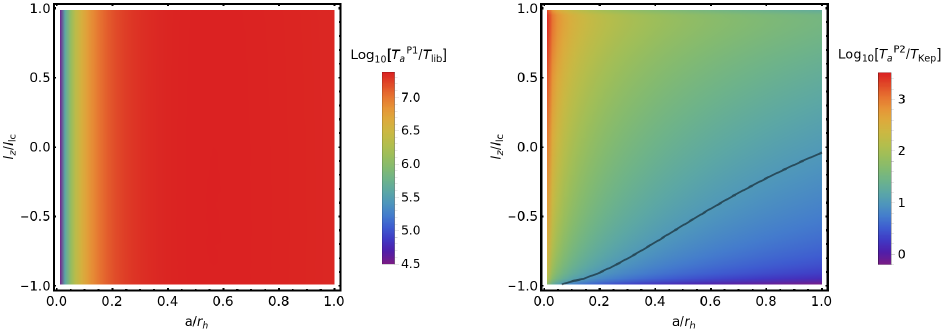}
    \caption{The ratio of migration to libration time $T_a/{\rm max}[\Tlib,\TKep]$ on the separatrix orbits, that eventually lead to collisionless TDEs, is plotted in $\{a,\lz\}$-plane. The left panel presents the relevant ratio $T_a/\Tlib$ at the point P1 for which $\ell = \lmax$ ($g = 90^{\circ}$), representing roughly the edge of libration island. Right panel shows the ratio $T_a/\TKep$ at point P2 with $\ell = \llc$ corresponding to the final collisionless TDE orbit. The black contour represents the parameters for which $T_a/\TKep = 10$. Since $T_a \gg \Tlib$ at point P1 and $T_a \gtrsim 10 \TKep$ at point P2 for most of the parameter space, suggesting low significance of gas dissipation in formation of collisionless TDEs. We choose $\Mbh = 10^7 \Msun $, $\mu_0 = 0.1$, $\beta = 0$, $\gamma = 7/4$ for this calculation.     
    } 
    \label{fig_Ta_den_plots}
\end{figure*}

\begin{figure}
    \centering
    \includegraphics[width=0.4\textwidth]{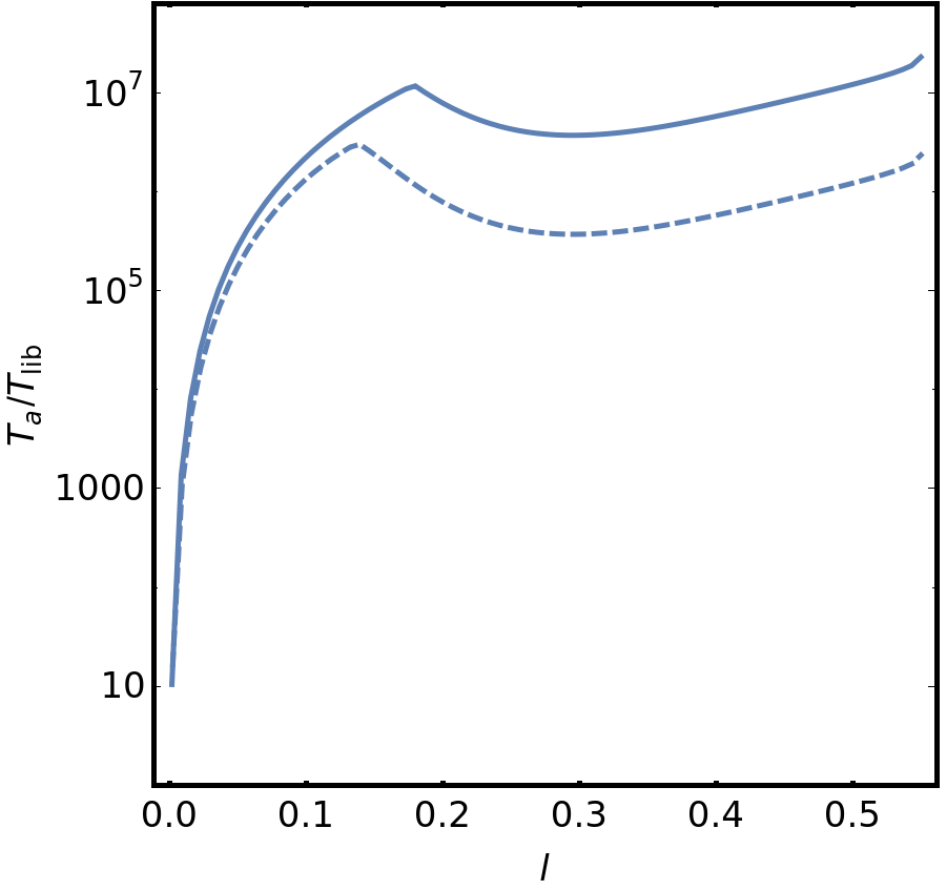}
    \caption{The ratio of migration to libration time $T_a/{\rm max}[\Tlib,\TKep]$ plotted along a separatrix orbit with $\ell \in [\llc, \lmax]$ for fixed $a = \rh$ and $\ell_z = 0$.  The sharp change in the behaviour close to $\ell \simeq 0.1-0.2$ is due to dominance of drag term for small $\ell$. This qualitative behaviour remains the same for different  values of $\ell_z/\llc$. During all phases $\ell$ of the libration cycle, this ratio remains greater than 10 suggesting only a limited significance of gas dissipation for collisionless TDE formation. The solid curve is for $M = 10^7 \Msun$ and the dashed one for $M = 10^6 \Msun$.   
    } 
    \label{fig_Ta_plot}
\end{figure}

We investigate the significance of dissipative effects in the entire loss wedge $\lz \in [-\llc,\llc]$ and $a \in (0,r_h]$ in figure~\ref{fig_Ta_den_plots}. We present $T_a/{\rm max}[\TKep,\Tlib]$ at the two extreme points of the separatrix orbit for each $\{a,\lz \}$. These extreme points on SO correspond to: (1) point P1 with $\ell = \lmax$ and $g = 90^{\circ}$ (left panel) which roughly defines the width of the libration island, and (2) point P2 with $\ell = \llc$ (right panel), the TDE loss cone. Here $\Tlib(\ell)$ is the timescale for a star at the libration phase $\ell$ to reach the loss cone $\llc$, and is computed numerically (see  appendix~\ref{app_actions} for details). Point P1 representing the edge of libration island, corresponds to orbits with high $i \sim 80^{\circ}-110^{\circ}$ and moderately high $e \sim 0.8-0.9$, which implies weak dissipation by disk leading to much longer $T_a > 10^4 \, \Tlib(\lmax)$. During the libration cycle, a star spends most time near high $\ell$ phases close to P1. Hence, the migration time $T_a$ evaluated at P1 is the most relevant one. P2 corresponds to loss cone orbits, %final collisionless TDE orbits, 
which are highly eccentric ($e \sim 0.999 - 0.9999$) with line of apses closely aligned with disk. %, along with a sinusoidal distribution in $i$ (from \S~\ref{sec_TDE_orbits}).
For most $\{a , \lz\}$, $T_a/\TKep \gtrsim 10$ (above the black contour). But for retrograde orbits closely aligned with disk, $T_a$ become shorter than $\TKep$. This apparently suggests a slight preference of collisionless TDEs towards prograde orbits with respect to disk rotation. Our simple analysis suggests that most stellar orbits captured in the libration island, would librate freely to ultimately become collisionless TDEs, without much interference by gas dissipation. Notably, our treatment of evaluation of $\Fdis$ only at nodes is not valid for orbits aligned closely with disk plane $i \sim 0^{\circ}, 180^{\circ}$, but such orbits would constitute very small fraction of collisionless TDEs. In figure~\ref{fig_Ta_plot}, we check this time ratio for all phases $\ell \in [\llc,\lmax]$ of a SO, which further reinforces the above conclusions on the role of gas dissipation in formation of collisionless TDEs.

\subsection{Mathematical details}

Here we give the remaining mathematical expressions needed to evaluate the migration timescale $T_a$ for a Keplerian orbit with given elements $\{a, e, i, g \}$. The stellar orbit intersects the mid-plane of anti-clockwise rotating disk at nodal points $N_1$ (ascending, $\phi_p = 0^{\circ}$) and $N_2$ (descending, $\phi_p = 180^{\circ}$) with the radial distance $r_n$: 
\beq 
r_n = \frac{a (1-e^2)}{1 + e \cos{(\phi_p - g)}} \,.
\eeq 

At a given node, velocity of star $\vstarvec$ has the magnitude:
\beq 
\vstar = \sqrt{G \Mbh \bigg( \frac{2}{r_n} - \frac{1}{a} \bigg)}
\eeq 
and the following velocity components along the unit vectors $\{\rcap, \phicap_p , \zcap_p \}$ for the coordinate system (defined above) aligned with stellar orbital plane: 
\beq 
\begin{split} 
& \vstarr = \frac{2 \pi a }{\TKep(a)\sqrt{1-e^2}} e \sin{(\phi_p - g)} \\[1ex] 
& \vstarphip =  \frac{2 \pi a }{\TKep(a)\sqrt{1-e^2}} \big( 1 + e \cos{(\phi_p - g)}  \big) \\[1ex]
& \vstarzp = 0  
\end{split} 
\eeq 
Also, the magnitude of the stellar velocity component (at a node) perpendicular to the disk plane is $|\vstarz| = \vstarphip \sin{i}$.  

The gas velocity $\vgvec$ at a given node has the magnitude $\vg = \sqrt{G \Mbh/r_n}$, with the velocity components: 
\beq 
\vgr = 0 \qquad ,  \qquad \vgphip = \vg \cos{i} \qquad , \qquad \vgzp  = \vg \sin{i} \cos{(\pi - \phi_p)}
\eeq 

The relative velocity $\vrelvec = \vstarvec - \vgvec$ of star with respect to gas at a node, has the magnitude: 
\beq 
\vrel = \sqrt{ \vstar^2 +\vg^2 - 2 \, \vg\,  {\vstarphip}\, \cos{i} }
\eeq

 \end{document}